\documentclass[conference]{IEEEtran}
\IEEEoverridecommandlockouts
\usepackage{cite}
\usepackage{amsmath,amssymb,amsfonts}
\usepackage{algorithmic}
\usepackage{graphicx}
\usepackage{textcomp}
\usepackage{xcolor}

\usepackage{subcaption}

\def\BibTeX{{\rm B\kern-.05em{\sc i\kern-.025em b}\kern-.08em
    T\kern-.1667em\lower.7ex\hbox{E}\kern-.125emX}}

\begin{document}

\title{An Evaluation of Three Distance Measurement Technologies for Flying Light Specks*\\
\thanks{This research was supported in part by the National Science Foundation grant IIS-2232382.}
}

\author{\IEEEauthorblockN{Trung Phan, Hamed Alimohammadzadeh, Heather Culbertson, Shahram Ghandeharizadeh}
\IEEEauthorblockA{\textit{Computer Science Department} \\
\textit{University of Southern California}\\
Los Angeles, CA, USA \\
\{taphan,halimoha,hculbert,shahram\}@usc.edu}


}

\maketitle

\begin{abstract}
This study evaluates the accuracy of three different types of time-of-flight sensors to measure distance.
We envision the possible use of these sensors to localize swarms of flying light specks (FLSs) to illuminate objects and avatars of a metaverse.
An FLS is a miniature-sized drone configured with RGB light sources. 
It is unable to illuminate a point cloud by itself.
However, 
the inter-FLS relationship effect of an organizational framework will compensate for the simplicity of each individual FLS, enabling a swarm of cooperating FLSs to illuminate complex shapes and render haptic interactions.
Distance between FLSs is an important criterion of the inter-FLS relationship.
We consider sensors that use radio frequency (UWB), infrared light (IR), and sound (ultrasonic) to quantify this metric.
Obtained results show only one sensor is able to measure distances as small as 1 cm with a high accuracy.
A sensor may require 
a calibration process that impacts its accuracy in measuring distance.
\end{abstract}
\section{Introduction}\label{sec:intro}
\begin{figure}
\centering
\includegraphics[width=0.9\columnwidth]{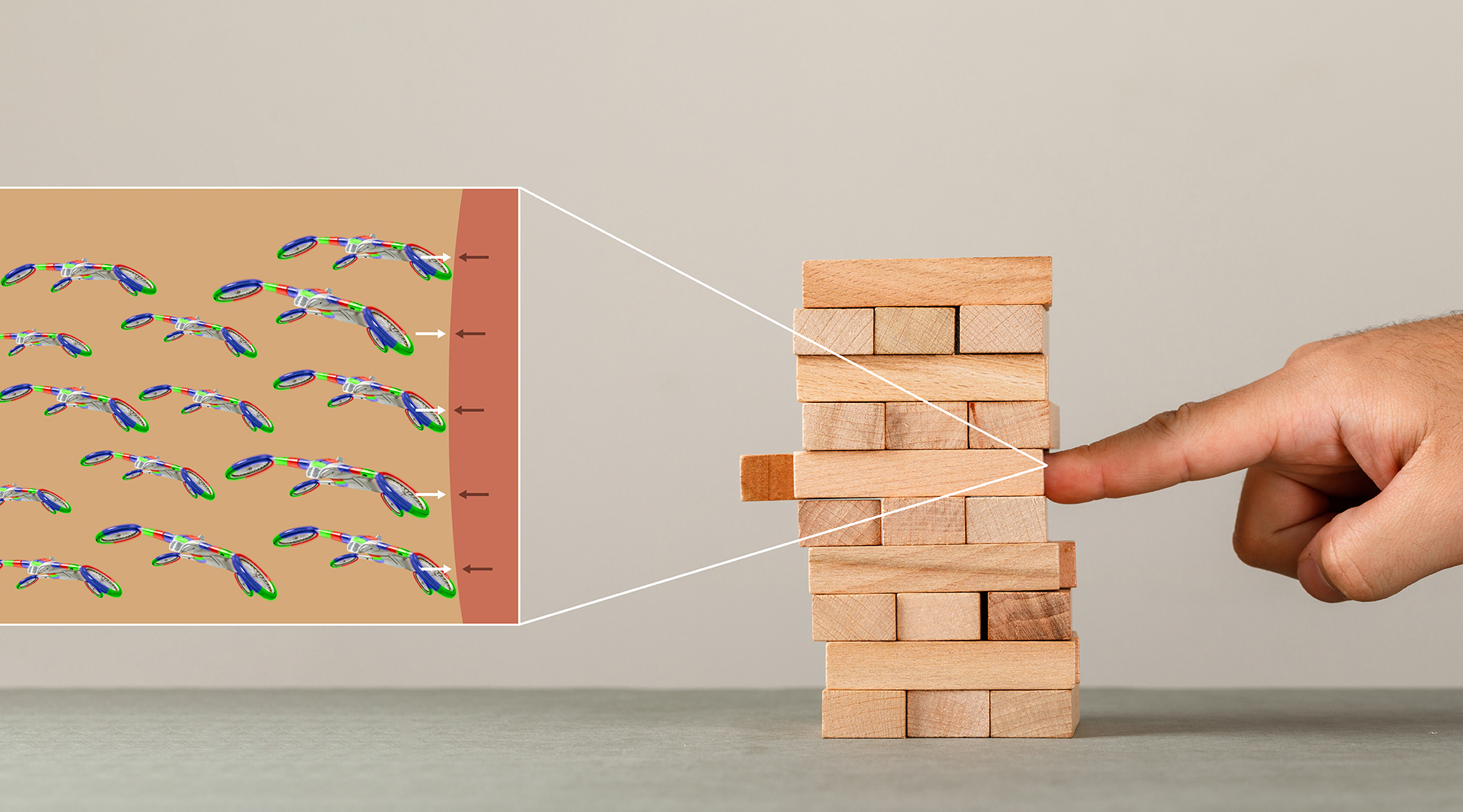}\hfill
\caption{FLSs rendering the game Jenga.}
\label{fig:janga}
\end{figure}

A Flying Light Speck, FLS, is a miniature sized drone configured with RGB light sources~\cite{shahram2021,shahram2022b,mmsys2023}.
Swarms of FLS may illuminate 3D objects and avatars in the metaverse
in a fixed 3D volume, an FLS display~\cite{shahram2022}.
These objects may provide users with haptic feedback, both kinesthetic (force) and tactile (skin-based).
Figure~\ref{fig:janga} shows FLSs illuminating the bricks of the game Jenga.
As the user pushes an illuminated Jenga brick, the FLSs will push back to provide kinesthetic feedback.
This force will be consistent with the surface friction between the brick and those below, above, and adjacent to it.

Both shape illuminations and kinesthetic haptic interactions require FLSs to localize.
Consider each in turn.  
To illuminate a shape, a swarm of FLSs must preserve a pre-specified relative distance between one another.
The pre-specified distance is defined by the points in a point cloud where each point is assigned to a different FLS~\cite{shahram2022}.
With kinesthetic feedback, when a user pushes on a swarm of FLSs, each FLS will compute its displaced distance and the speed of its displacement.
This information will be aggregated to quantify the force exerted by a user.
It will enable the swarm to exhibit the appropriate behavior (e.g., bricks at the top tower falling due to excessive force applied too fast) or generate force consistent with the user's expectation (e.g., the brick slides out gently while providing the sensation of surface resistance).

Measuring small distances accurately is fundamental to localizing FLSs.
The ideal sensor should exhibit the following features.
First, an FLS should be able to measure its distance to another FLS at a granularity that is a function of both its size and important physical parameters.
A physical parameter is downwash,
a region of instability caused by the flight of one FLS that adversely impacts other FLSs entering this region~\cite{downwash1,dcad2019,preiss2017,downwash3,Ferrera2018Decentralized3C,planning2019}, e.g., loss of control or unpredictable behavior.
FLSs must be able to measure distance at sufficient granularity to enable them to avoid each other's downwash.

Second, the measured distances should be highly accurate.
Error impacts both the quality of an illumination and how effectively FLSs render kinesthetic feedback. 
We define the tolerable error as a function of FLS size and important physical characteristics, e.g., 1\% of the length of an FLS, 1\% of the length/height/width of the downwash region.


\begin{small}
\begin{table}[b]
\caption{Three time-of-flight sensor types, their representatives, and advertised specifications:  Ultra Wide Band (UWB) (Decawave, DW1000)~\cite{dw1000}, Infrared (IR) (SparkFun, VL53L4CD~\cite{sparkfun} and Sharp, GP2Y0A51SK0F~\cite{sharpir}), and Ultrasonic (US) (ELEGOO, HC-SR04)~\cite{hcsr04}.  
}\label{tbl:sensors}
\begin{tabular}{|c|c|c|c|c|c|}
\hline
\hline
 & Calibrate & Min-Max  & Error  & Ran- & Price/ \\
  & Required & Distance & & ging &  Sensor \\
\hline
\hline
UWB~\cite{dw1000} & Yes & 10-29K cm & $\leq$10 cm & 2-way & \$40 \\
IR$_{Cal}$~\cite{sparkfun} & Yes & 0-130 cm & $\leq$0.1 cm & 1-way & \$20 \\
IR$_{NoC}$~\cite{sharpir} & No & 2-15 cm & $\leq$0.1 cm & 1-way & \$12 \\
US~\cite{hcsr04} & No & 2-400 cm & $\leq$0.3 cm & 1-way & \$1.80 \\

\hline
\hline

\end{tabular}
\end{table}
\end{small}

The application requirements will also be an important factor in dictating both the granularity of measured distance and its tolerable delay.
We anticipate these requirements to start simple and evolve to define next-generation devices due to the novelty of FLS displays.

In this study, we evaluate four sensors of three different technologies to measure distance between FLSs, see Table~\ref{tbl:sensors}.
Our objective is to accurately measure distances as small as 1 cm.
Our experiments are conducted as follows.
After calibrating a sensor (if required), we place a drone a fixed distance $D$ away from the sensor and gather its reported distances for 60 seconds.
We partition these into non-overlapping windows of time, each with duration $\delta$ seconds.
For each window of time, we report the minimum, average, and maximum percentage error.
With a window of time consisting of $M$ samples $\{S_1, S_2, \dots, S_M\}$, we define the average percentage error as: $100 \times \frac{\sum_{i=1}^M{S_i-D}}{M}$.

\begin{figure}
\centering
\includegraphics[width=0.9\columnwidth]{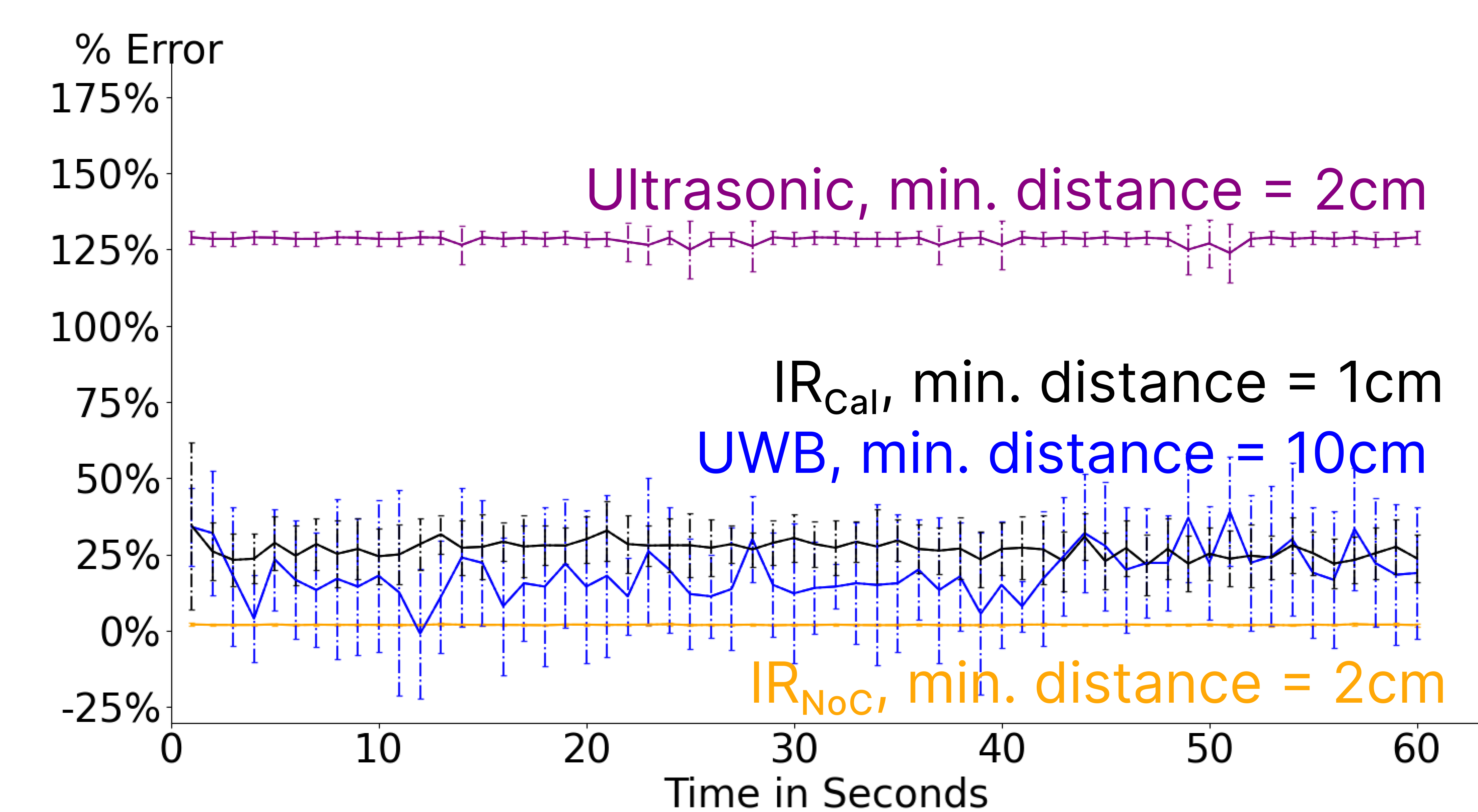}\hfill
\caption{A comparison of 4 sensors in measuring their minimum advertised distance.
We set the minimum at 1 cm with IR$_{Cal}$.  In these experiments, IR$_{Cal}$ was calibrated at 1 cm and UWB was calibrated at 10 cm.}
\label{fig:cmp}
\end{figure}

\noindent{\bf Summary of results:}
In our experiments, only one infrared sensor, IR$_{Cal}$, is able to measure distances as small as 1 cm with approximately 10-15\% error, 
see discussions of Figure~\ref{fig:infraredElevated}.
The minimum range of UWB, IR$_{NoC}$, and ultrasonic is greater than 1 cm, see Table~\ref{tbl:sensors}.
Hence, they are unable to measure 1 cm accurately.
Ultrasonic shows a high margin of error (125\%) when measuring its minimum distance, 2 cm, see Figure~\ref{fig:cmp}.
The margin of error is lower with UWB when measuring its minimum distance of 10 cm.
However, it is still significant at 20-30\%.

Both IR$_{Cal}$ and UWB measure a distance more accurately when calibrated for that distance.
Their percentage error is close to zero with distances of 20 cm and 100 cm.
Ultrasonic is equally accurate with a distance of 20 cm and does not require calibration.
It is possible to calibrate IR$_{Cal}$ and UWB at one distance and measure a different distance.
This increases their margin of error.

The primary {\bf contribution} of this paper is to quantify the percentage error observed with the sensors of Table~\ref{tbl:sensors} for specific experimental setups.
The rest of this paper is organized as follows.
Section~\ref{sec:related} presents the related research.
Sections~\ref{sec:uwb},~\ref{sec:IR}, and~\ref{sec:us} present the sensors in turn, our experimental setups, and obtained results.
Brief conclusions and future works are presented in Section~\ref{sec:conc}.

\section{Related Work}\label{sec:related}
To the best of our knowledge, the quantitative analysis of the four sensors considered in this study is novel and has not been published elsewhere.
Many studies have considered techniques to localize robots~\cite{cave2011,loc2021,radioslam2023}.
They have investigated environments with either the robot as the only
moving object (static) or other moving objects (dynamic)~\cite{dynamic2023,cave2011,goswami2023}.
Other studies consider simultaneous localization and mapping (SLAM) of an environment, e.g., a coal mine~\cite{cave2011,cave2022,cave2023}.
Our objective to measure distance between FLSs is similar because we want to localize FLSs relative to one another to illuminate objects and provide haptic interactions. 
At the same time, our environment is different along several dimensions.
First, the dimensions of the 3D display volume will be known in advance and used indoors. 
Second, a large number of miniature-sized FLSs will cooperate in this volume, thousands if not millions.
The novelty is for the FLSs to maintain a relative distance from one another (illumination) and to measure their displacement due to haptic user interactions.

Centralized indoor techniques such as optical motion capture systems merge images from fixed cameras positioned around a display volume to localize quadrotors and drones~\cite{opticalpositioning1,opticalpositioning2,preiss2017}.
These systems, e.g., Vicon, are highly accurate.
They require a unique marker arrangement for each drone and have enabled control and navigation of swarms of tens of drones.
It may be difficult (if not impossible) to form unique markers for more than tens of small drones measuring tens of millimeters diagonally~\cite{preiss2017}.
Hence their scalability is limited.
Moreover, these systems require broadcast at high frequency from a central computer to each drone~\cite{weinstein2018}.
This centralized communication channel is a single point of system failure.
It has a latency of several (7) milliseconds~\cite{preiss2017} and its bandwidth will constrain the robustness and FLS swarm size.
We envision FLSs with sensors to quantify distance in a decentralized manner. 
\begin{figure}[h]
\centering
\begin{subfigure}[t]{1\linewidth}
  \centering
  \includegraphics[width=\linewidth]{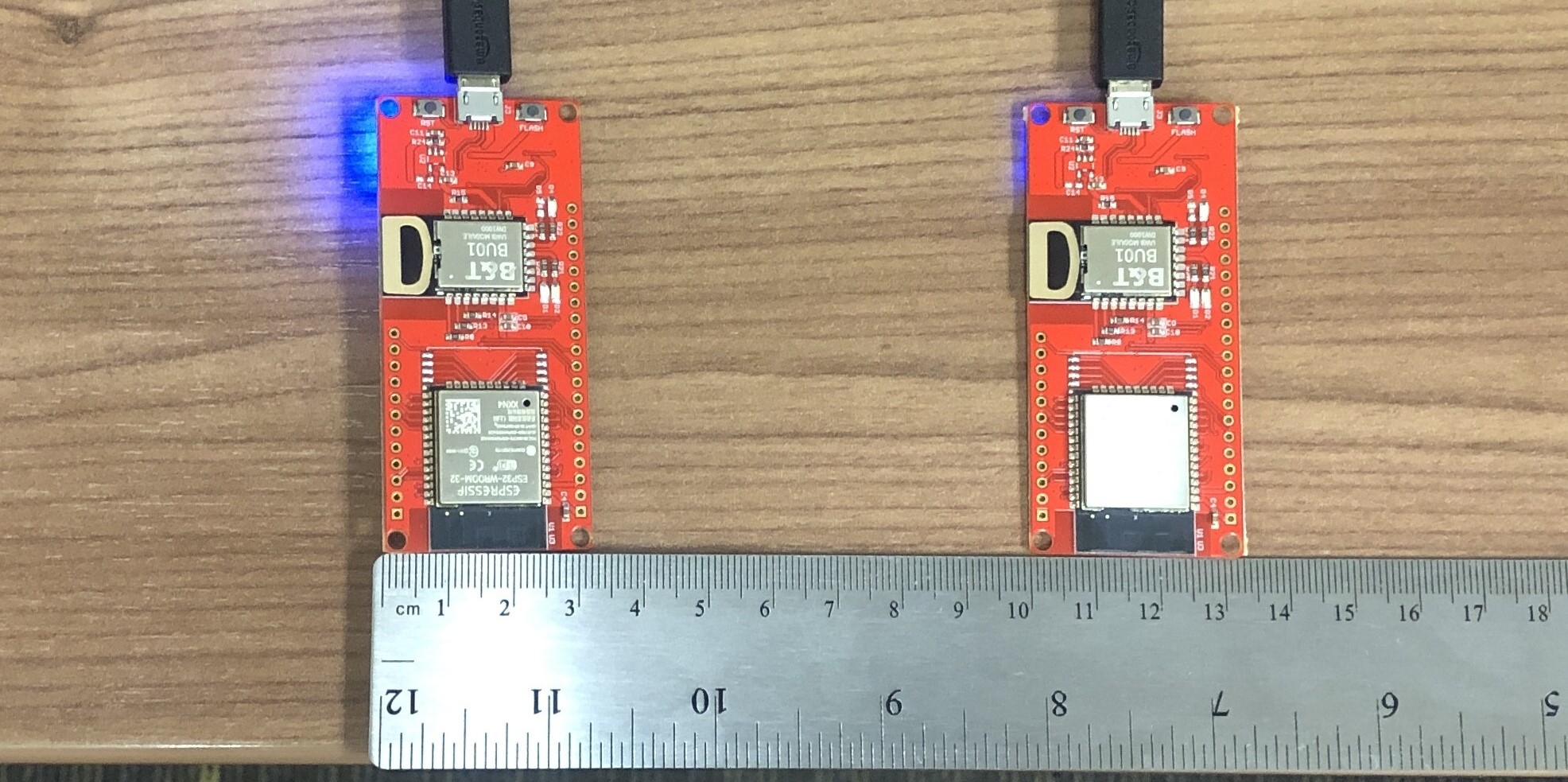}
  \caption{Tabletop.}
\end{subfigure}
\quad

\begin{subfigure}[t]{1\linewidth}
  \centering
  \includegraphics[width=\linewidth]{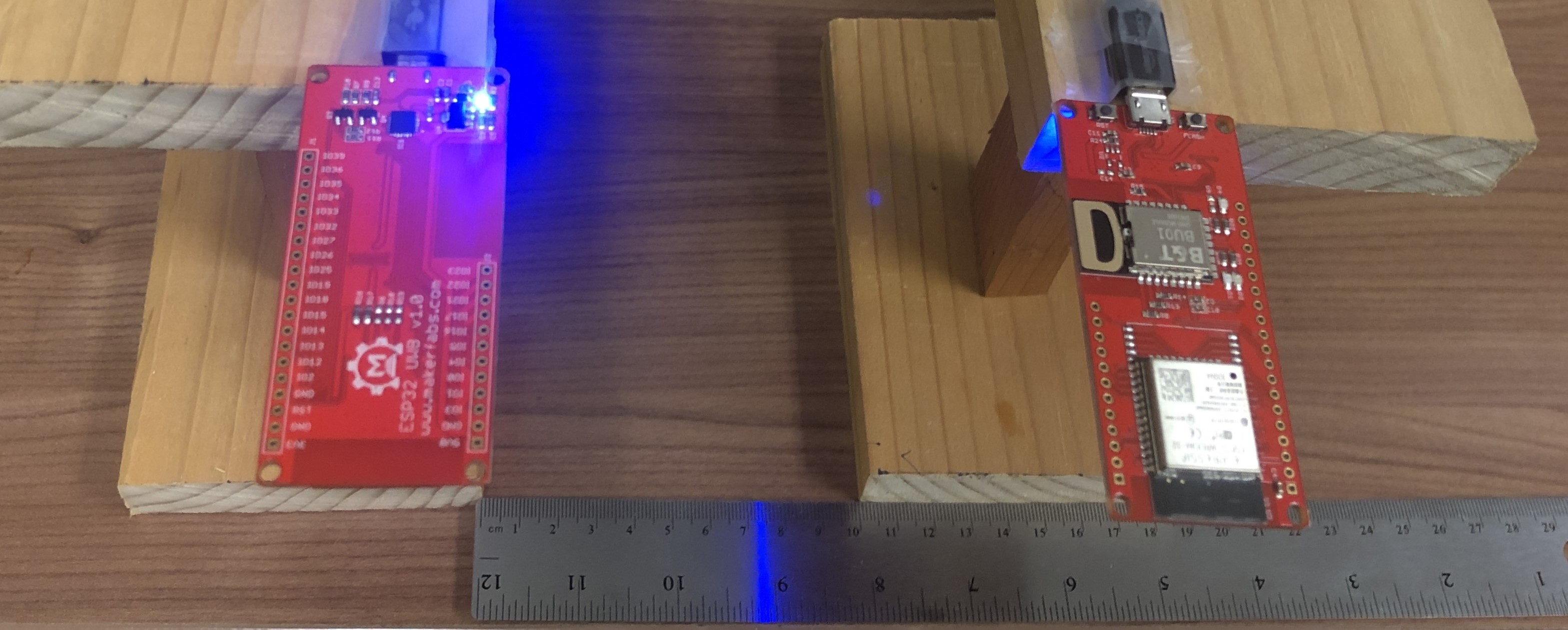}
  \caption{Suspended horizontally.}
\end{subfigure}

\quad

\begin{subfigure}[t]{1\linewidth}
  \centering
  \includegraphics[width=\linewidth]{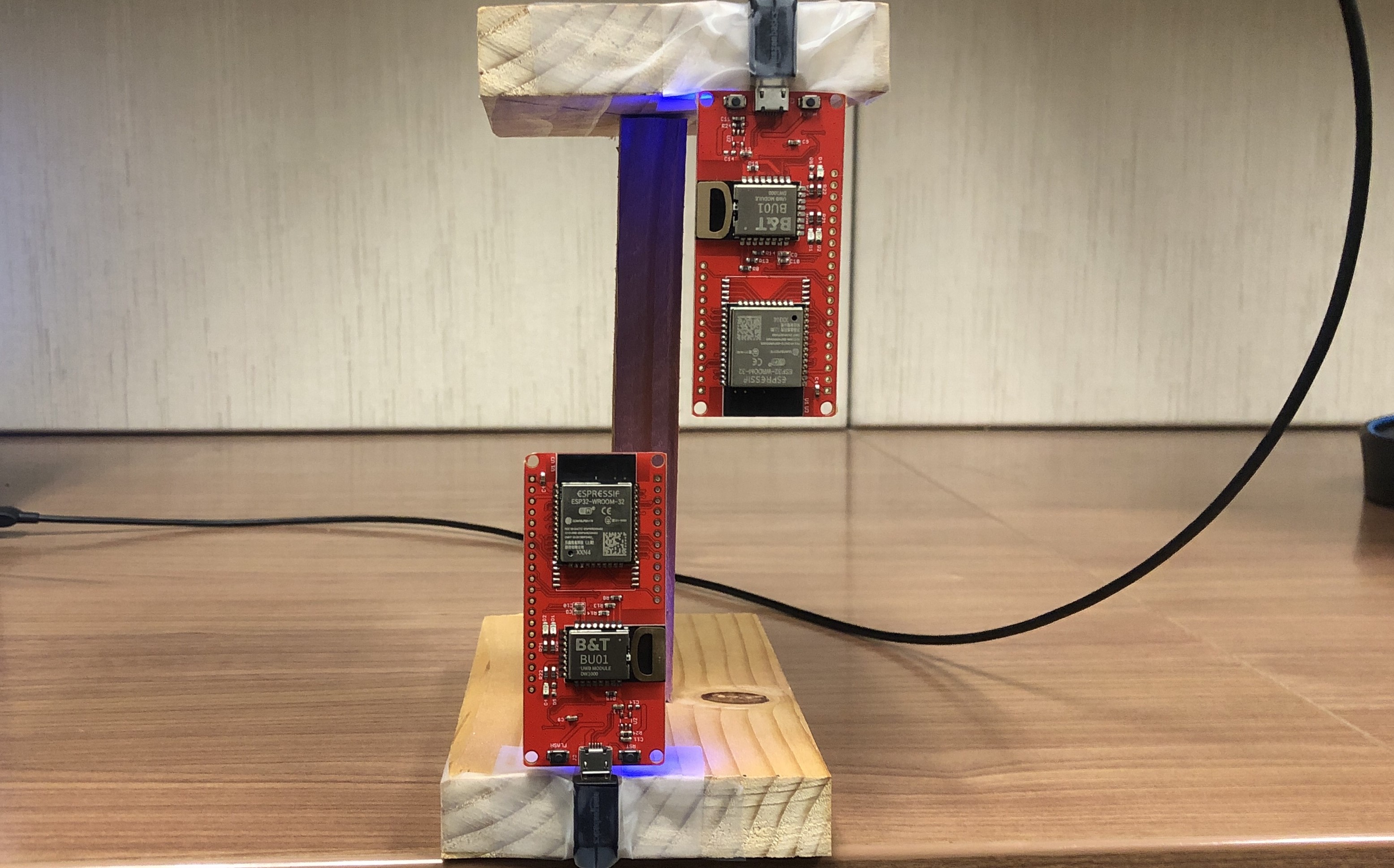}
  \caption{Suspended diagonally.}
\end{subfigure}

\caption{UWB experimental setups.}
\label{fig:uwbsetup}
\end{figure}

\section{Ultra Wide Band, UWB}\label{sec:uwb}

UWB is a radio frequency technology operating in a wide bandwidth~\cite{uwb2001}. 
Its short pulse duration enables determining the time-of-flight of the radio signal with fine granularity. Different techniques can be used for ranging and localization. With a two-way ranging technique, the distance between two devices can be measured as follows:
\begin{equation}
    d = \frac{v_s \times (t_{rt}-t_{d})}{2}
\end{equation}
Where $v_s$ is the speed of the radio signal, $t_{rt}$ is the round trip time of the signal from source to destination and from the destination back to the source, and $t_d$ is the total time spent to process the signal in the source and destination.
Theoretically, UWB's short-duration pulses, in the order of nanoseconds, can provide centimeter accuracy in distance measurements~\cite{dw1000}.

This study uses the DW1000, a single chip CMOS radio transceiver.
It is compliant with the IEEE 802.15.4-1011 ultra-wideband (UWB) standard.  
It facilitates proximity detection to an accuracy of $\pm$ 10 cm using two-way ranging.

The provider of DW1000, Decawave, publishes a device driver for DW1000.
This includes a set of APIs to initialize, configure, and measure distance using a pair of DW1000 cards.
We used this software running on an Acer Swift 3 Windows laptop to make our measurements.

Figure~\ref{fig:uwbsetup} shows three alternative experimental setups for the UWB cards.
The suspended cards, both horizontal and diagonal, provide the most accurate readings.

\begin{figure*}[h]
\centering
\begin{subfigure}[t]{0.31\linewidth}
  \centering
  \includegraphics[width=\linewidth]{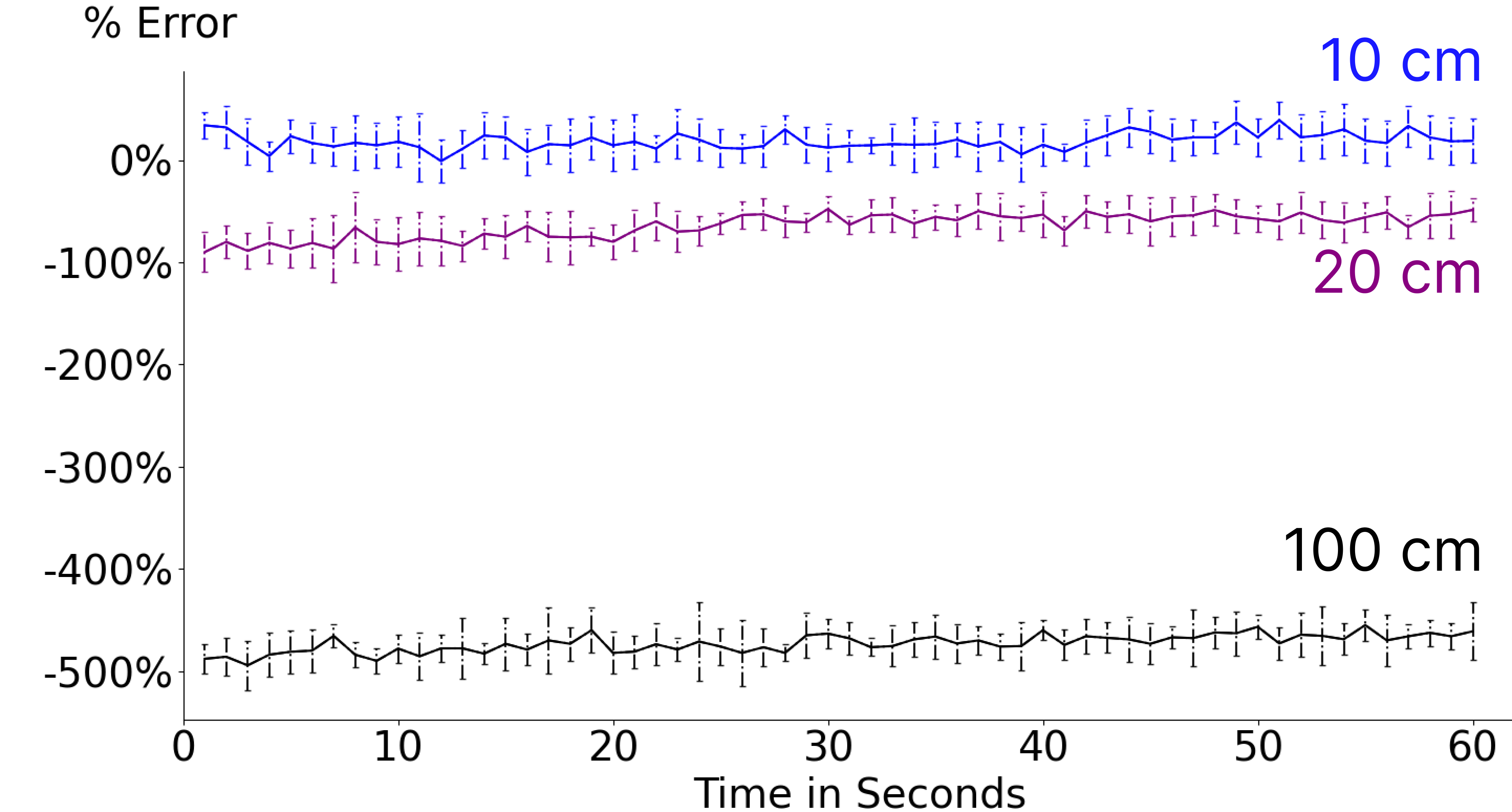}
  \caption{Percentage error to measure 10 cm distances when calibrated at 10, 20, and 100 cm.}\label{fig:uwbvarcalib}
\end{subfigure}
\quad
\begin{subfigure}[t]{0.31\linewidth}
  \centering
  \includegraphics[width=\linewidth]{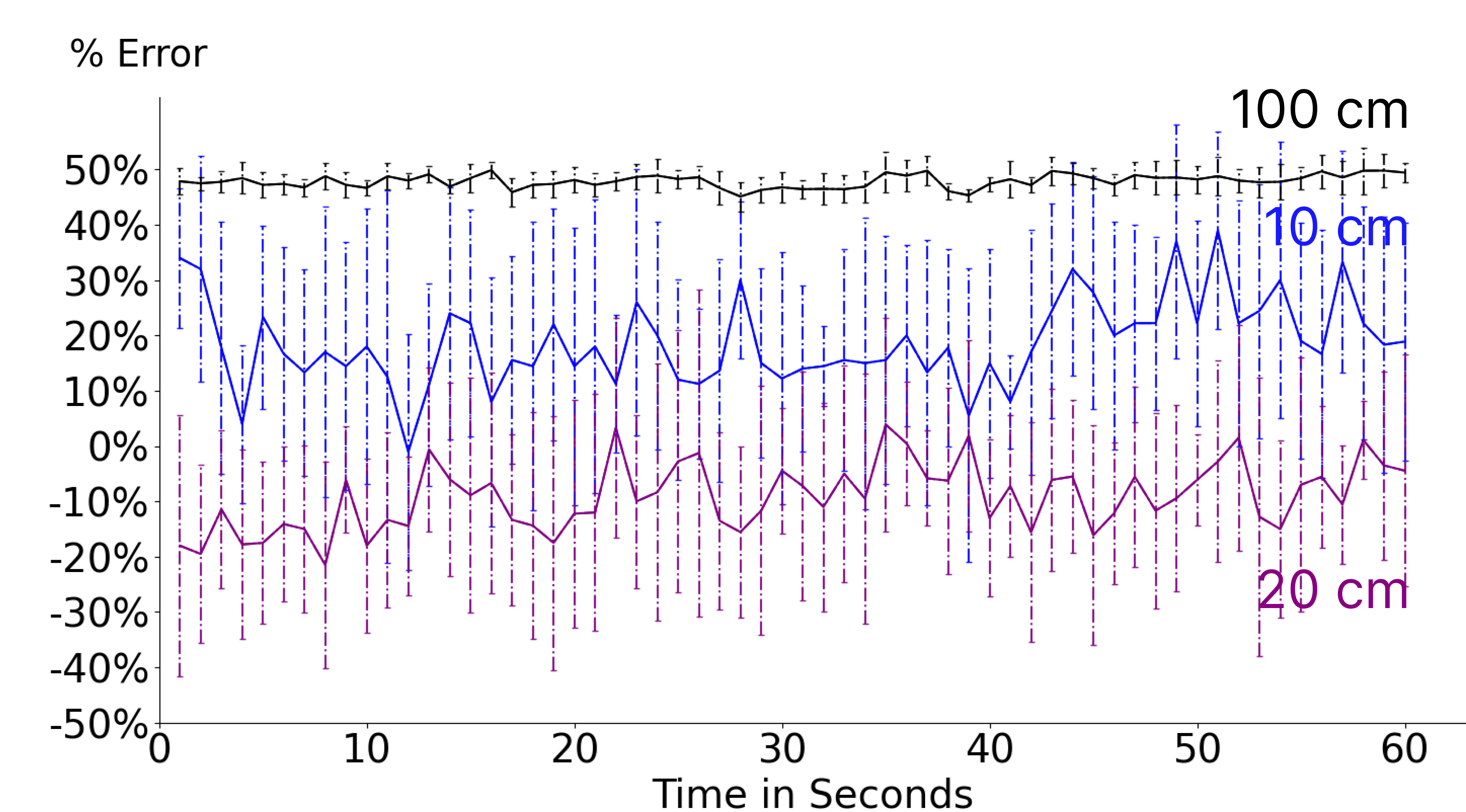}
  \caption{Percentage error in measuring 10, 20, and 100 cm distances with a 10 cm calibration.}\label{fig:uwb10cmcalib}
\end{subfigure}
\quad
\begin{subfigure}[t]{0.31\linewidth}
  \centering
  \includegraphics[width=\linewidth]{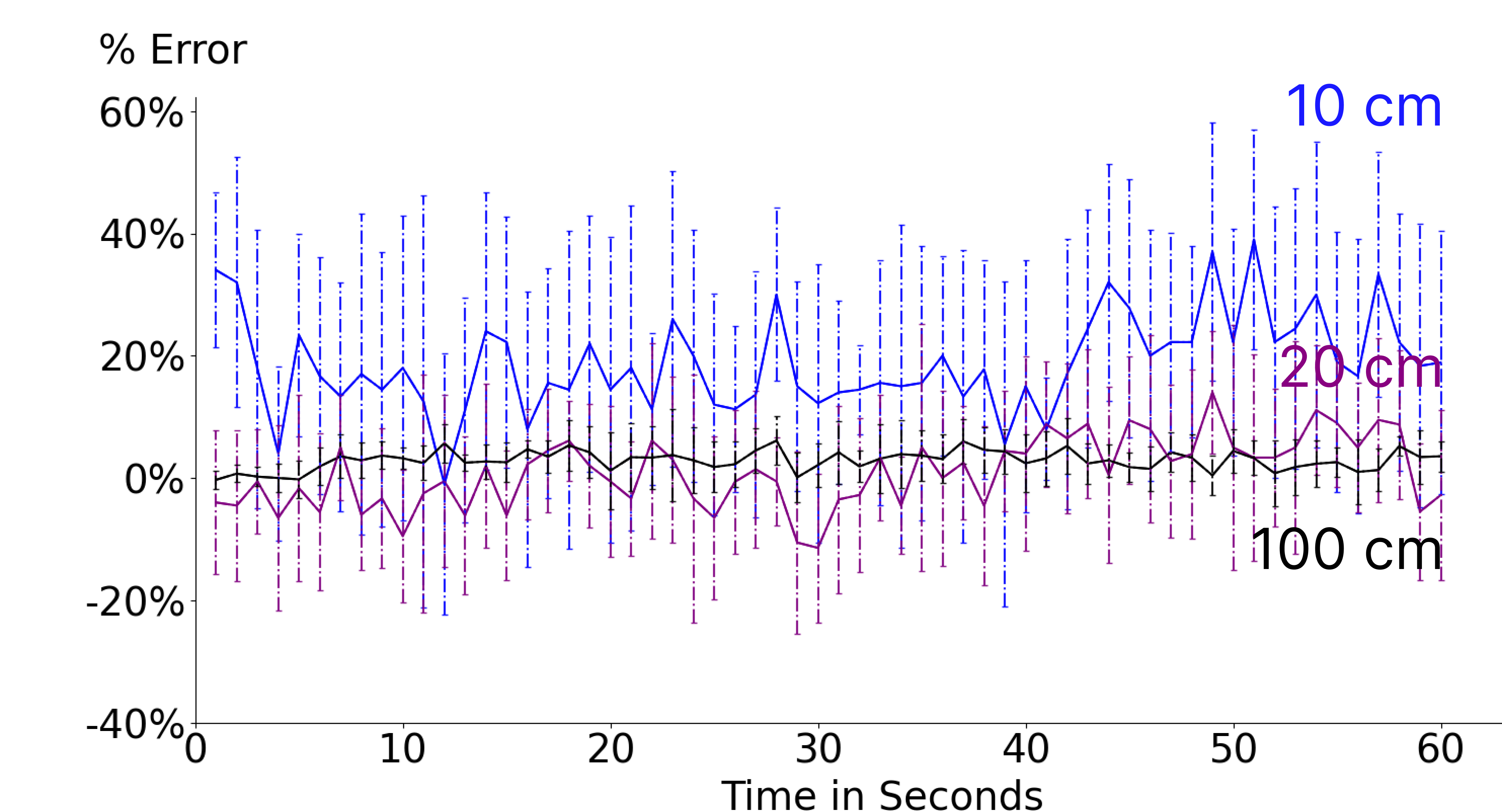}
  \caption{Percentage error in measuring 10 cm with 10 cm calibration, 20 cm with 20 cm calibration, and 100 cm with 100 cm calibration.}\label{fig:uwbdiffcalib}
\end{subfigure}
\caption{UWB quantifying different distances using alternative calibrations.}
\label{fig:uwb}
\end{figure*}

The accuracy of measured distances with UWB is impacted by its calibration process.
Figure~\ref{fig:uwbvarcalib} shows the percentage error in measuring distance between two cards 10 cm apart.
The cards were calibrated at 10, 20, and 100 cm.
The percentage error increases dramatically with higher calibrations. 
Figure~\ref{fig:uwb10cmcalib} shows the percentage error in making three different distance measurements with a calibration of 10 cm.
The highest percentage error is observed with a larger distance, i.e., 100 cm. 
If we change the calibration to 20 cm, the curve for 20 cm and 10 cm switch places, i.e., the measurements for 20 cm become more accurate.
However, the measurements for 100 cm remain as inaccurate.

Figure~\ref{fig:uwbdiffcalib} shows UWB is more accurate when calibrated for the distance it is intended to measure.
Its accuracy increases with longer distances.
Its average percentage error is close to 0\% when measuring 100 cm distances with UWB calibrated at 100 cm.

\noindent{\bf Related Work:}  There are studies that use the DW1000 with anchors placed at well known coordinates to localize drones~\cite{uwb2015,snaploc2019}.
Imperfect anchor clock synchronization~\cite{clockdrift22} are reported as the cause of inaccuracies exceeding 7.5 cm~\cite{uwb2015}.
SnapLoc~\cite{snaploc2019} employs wired communication between UWB anchors to enhance localization accuracy.
It reports longer distances may be measured accurately with a higher probability, e.g., 33.7 cm distances are measured accurately with a 90\% probability and 60 cm distances are measured accurately with a 99\% probability.
\section{Infrared}\label{sec:IR}

\begin{figure*}
\centering
\begin{subfigure}[t]{0.47\linewidth}
  \centering
  \includegraphics[width=\linewidth]{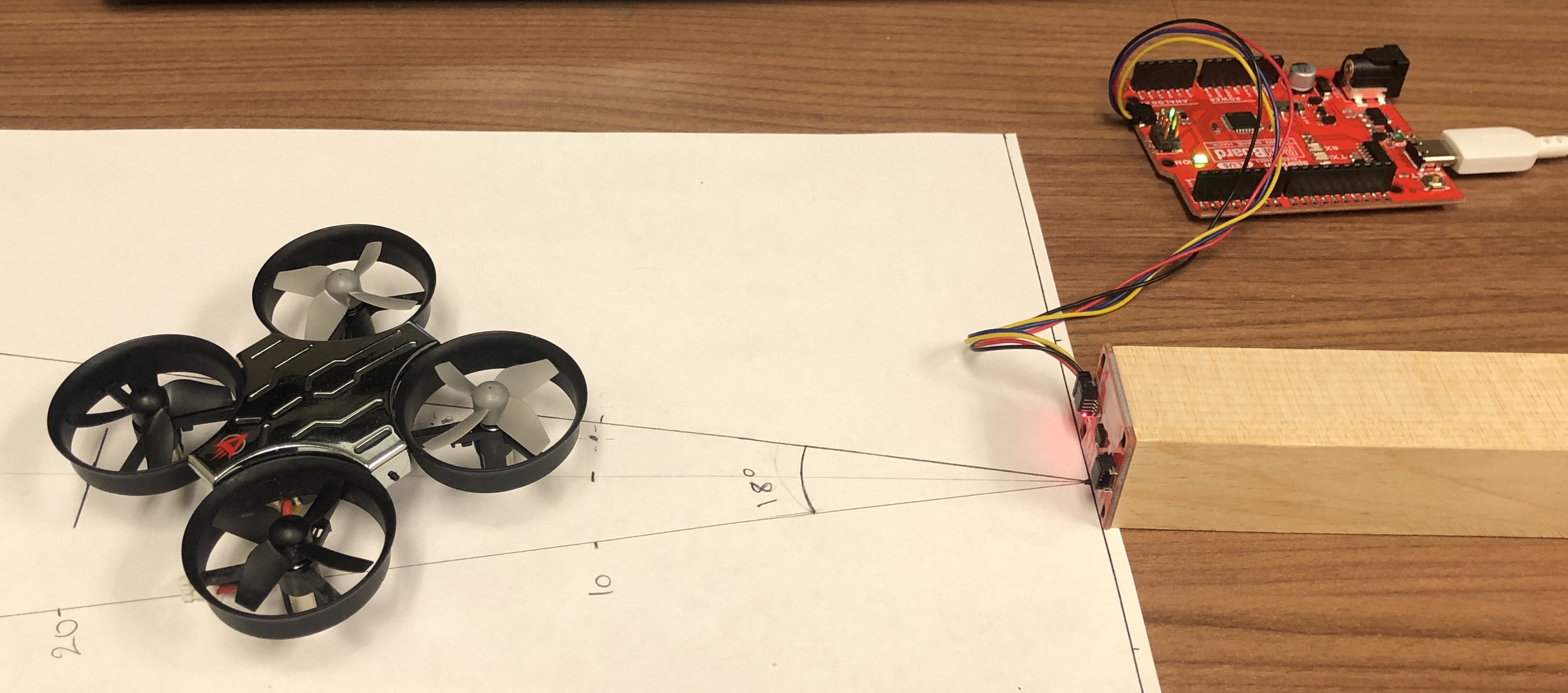}
  \caption{IR$_{Cal}$ sensor is touching the sheet.}
\end{subfigure}
\quad
\begin{subfigure}[t]{0.47\linewidth}
  \centering
  \includegraphics[width=\linewidth]{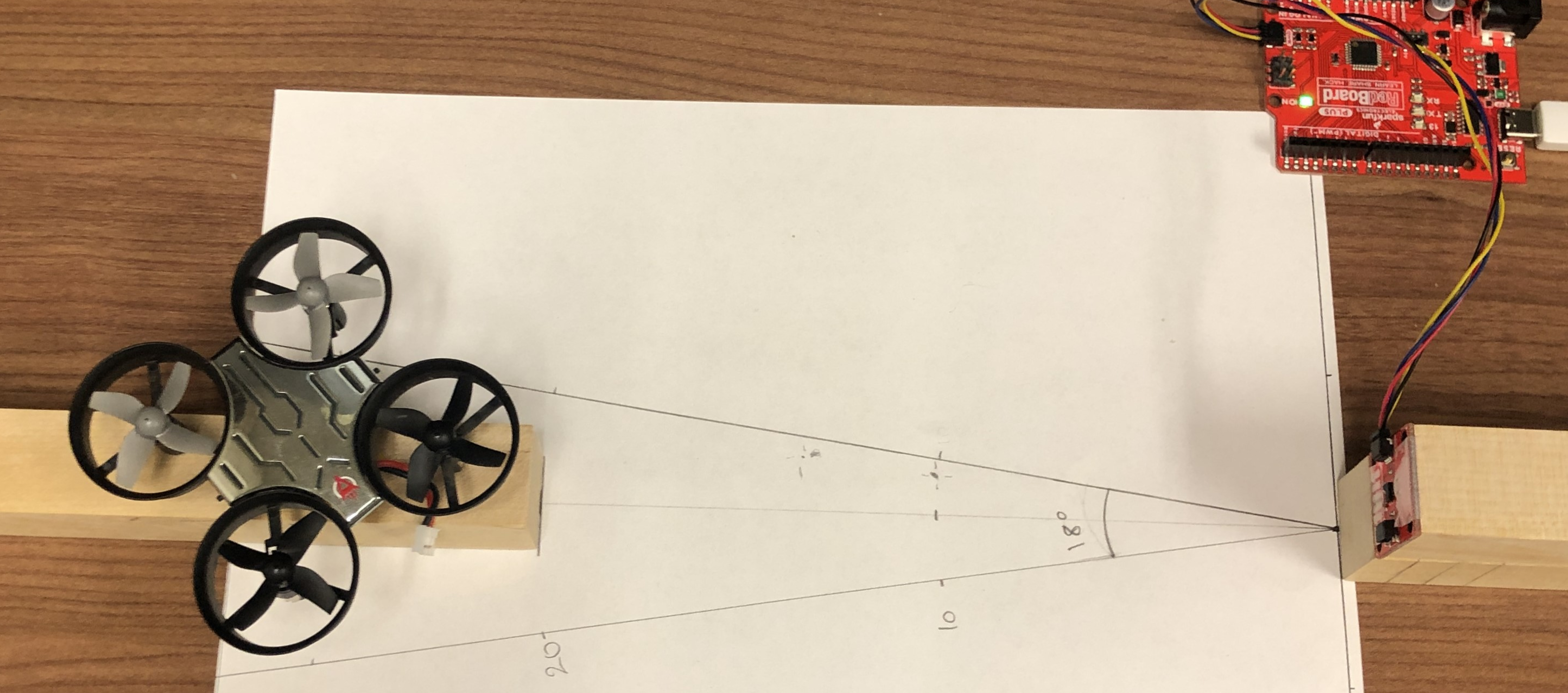}
  \caption{IR$_{Cal}$ sensor is elevated 3.7 cm.}
\end{subfigure}
\caption{Elevation impacts accuracy of measured distances with IR$_{Cal}$.}
\label{fig:infrared}
\end{figure*}


Infrared (IR) sensors are active sensors that emit a signal (IR light) and then detect the signal's return after it bounces off of the object. The sensor calculates the distance to the object using the time-of-flight (i.e., the amount of time recorded between emitting and detecting the light):
\begin{equation}
    d=\frac{c \times t}{2}
\end{equation}
where $c$ is the speed of light and $t$ is the recorded time.

IR sensors can sense closer objects and have a higher resolution than ultrasonic sensors, but they have an overall shorter range. A typical IR sensor has a range of 0--130 cm with a resolution of 1 mm~\cite{benet2002using}.
However, this range can be affected by the color of the object it is detecting. While color does not change the amount of time it takes for the light to travel, lighter colors reflect more light than darker colors. Therefore, the sensor will perform optimally when detecting white objects within a specific distance range. (i.e., reflecting the most light) and its worst distance range will be when detecting black objects (i.e., reflecting the least light). 

The wavelength of IR sensors (940 nm) is also significantly less than that of ultrasonic sensors, meaning that they can detect significantly smaller objects than ultrasonic sensors~\cite{benet2002using}. This parameter is very important in FLS systems. IR sensors also have a smaller FoV (18$^\circ$) than ultrasonic sensors, creating a higher spatial resolution.

One limitation of IR sensors is that their performance can be affected by external light in the environment. Significant IR light from the environment can saturate the sensor or cause incorrect distance readings. Therefore, these sensors are more accurate indoors than outdoors. Best practices with IR sensors includes shielding the sensor from external light to mitigate any errors.

We evaluate the 2 IR sensors shown in Table~\ref{tbl:irsensors}, the difference between them being that one requires calibration (IR$_{Cal}$) and the other does not (IR$_{NoC}$).

\begin{small}
\begin{table*}[htbp]
\begin{center}
\caption{Two Infrared (IR) sensors.  SparkFun, VL53L4CD~\cite{sparkfun}, and Sharp, GP2Y0A51SK0F~\cite{sharpir}.
}\label{tbl:irsensors}
\begin{tabular}{|c|c|c|c|c|c|c|}
\hline
\hline
 & Manufacturer & Calibration & Min-Max  & Error  & Ranging & Price/ \\
 & Model Number & Required & Distance & &  &  Sensor \\
\hline
\hline

IR$_{Cal}$ & SparkFun:VL53L4CD~\cite{sparkfun} & Yes & 0-130 cm & $\leq$0.1 cm & 1-way & \$20 \\
IR$_{NoC}$ & Sharp:GP2Y0A51SK0F~\cite{sharpir} & No & 2-15 cm & Not Available & 1-way & \$12 \\

\hline
\hline
\end{tabular}
\end{center}
\end{table*}
\end{small}

\subsection{IR$_{Cal}$, SparkFun:VL53L4CD}

\begin{figure*}
\centering
\begin{subfigure}[t]{0.31\linewidth}
  \centering
  \includegraphics[width=\linewidth]{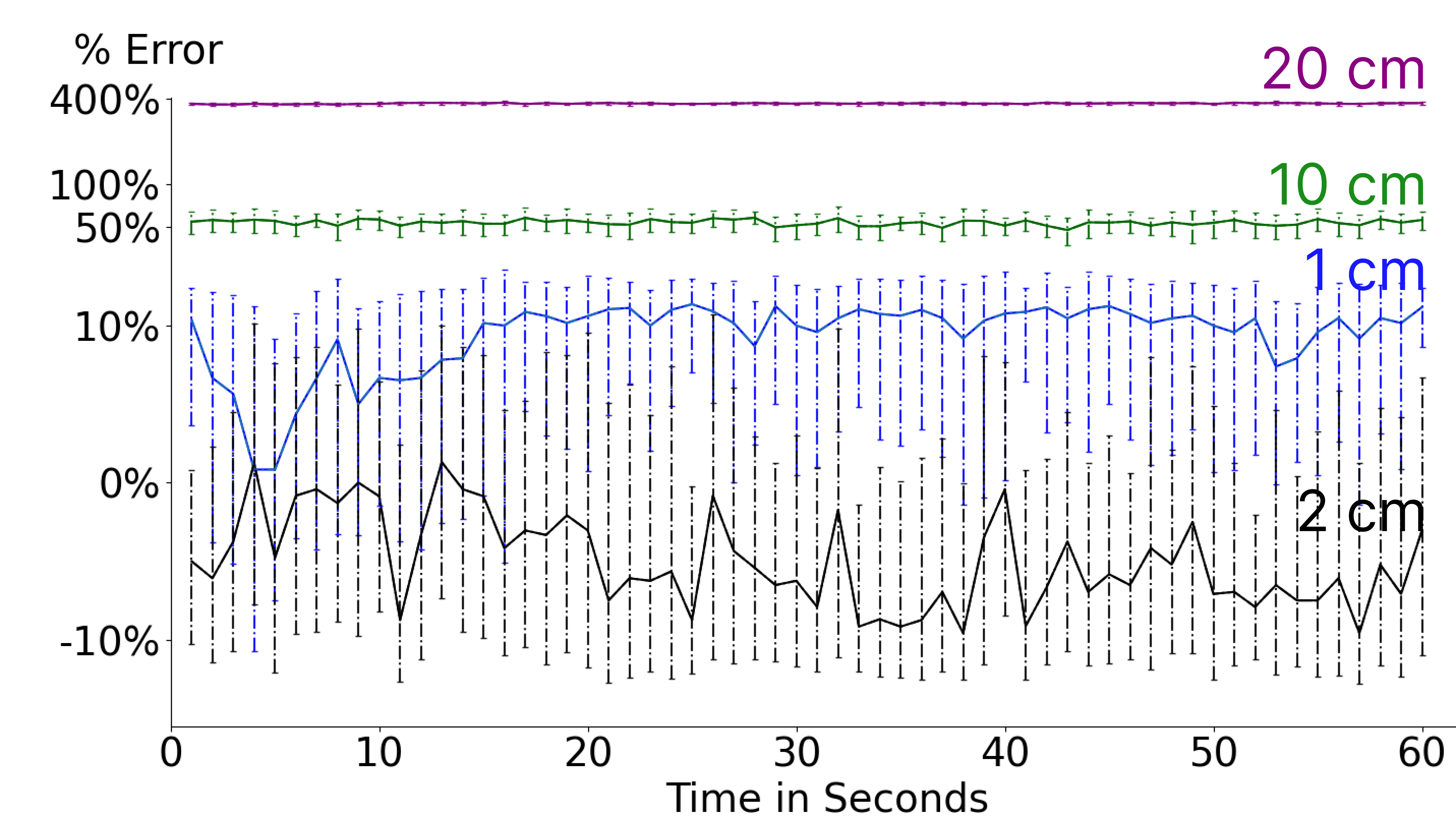}
  \caption{IR$_{Cal}$ sensor is touching the sheet.  Measuring different distances with a 1 cm calibration.}\label{fig:infraredTouching}
\end{subfigure}
\quad
\begin{subfigure}[t]{0.31\linewidth}
  \centering
  \includegraphics[width=\linewidth]{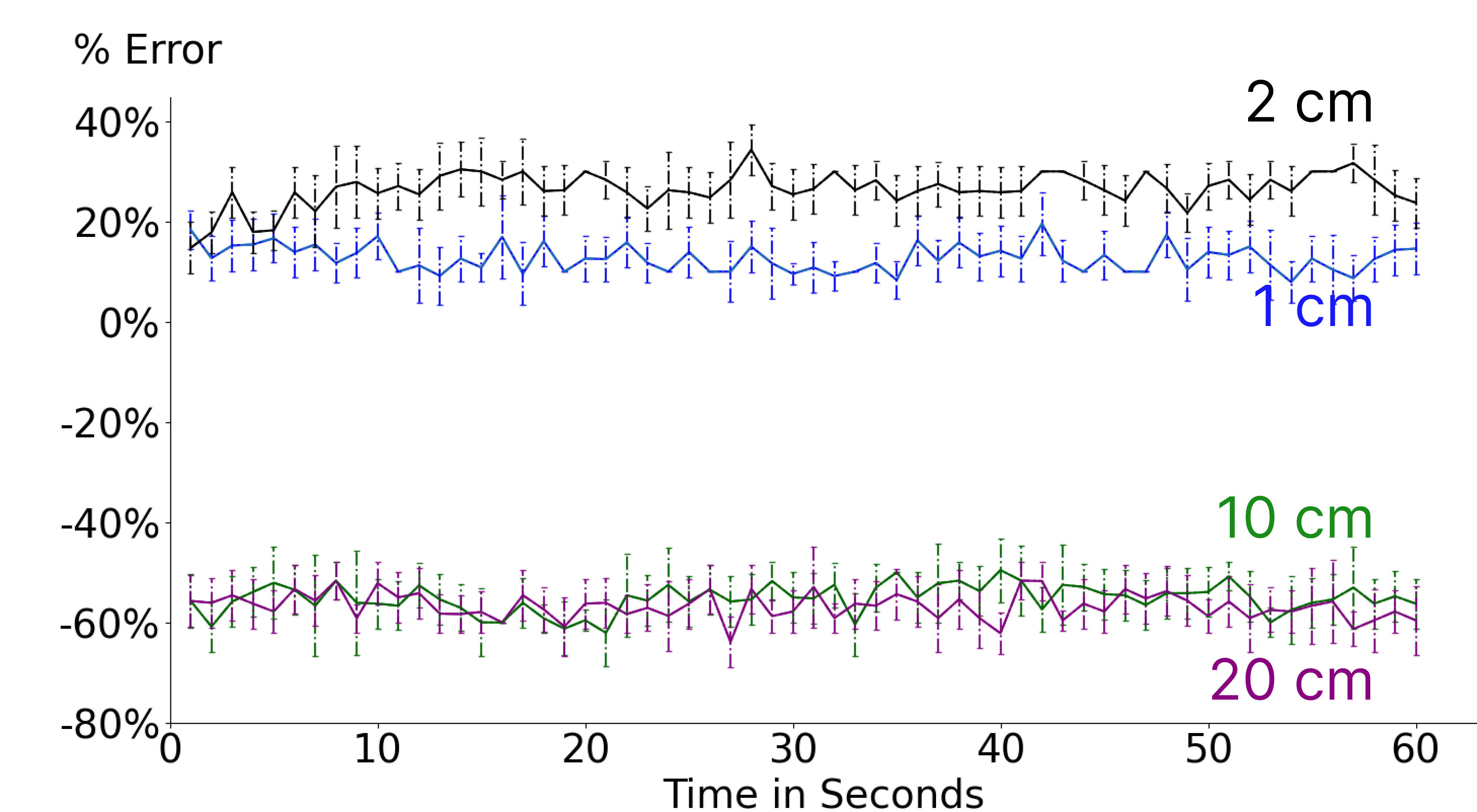}
  \caption{IR$_{Cal}$ sensor is elevated 3.7 cm.  Measuring different distances with a 1 cm calibration.}\label{fig:infraredElevated}
\end{subfigure}
\quad
\begin{subfigure}[t]{0.31\linewidth}
  \centering
  \includegraphics[width=\linewidth]{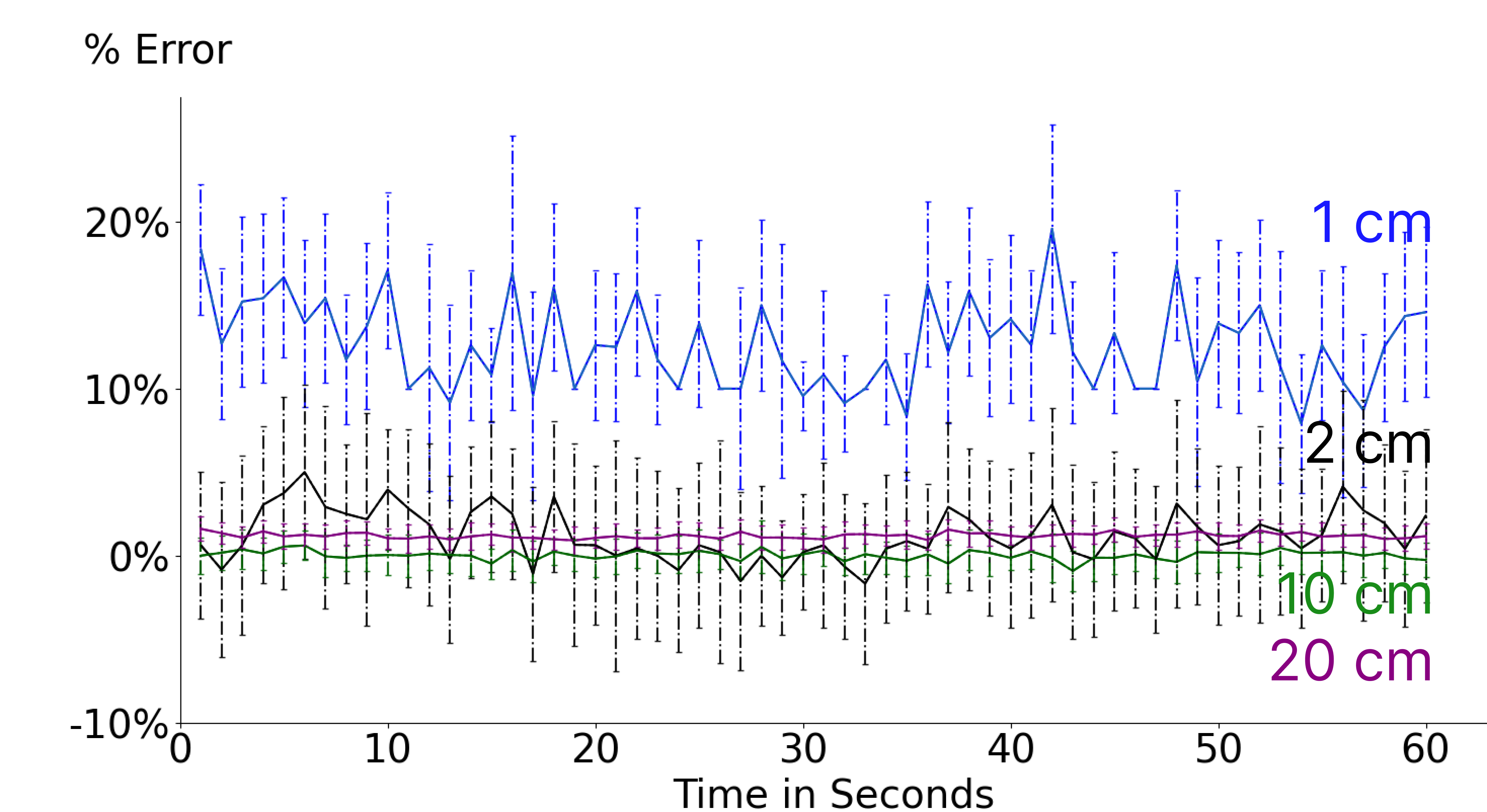}
  \caption{Percentage error in measuring 1 cm with 1 cm calibration, 2 cm with 2 cm calibration, 10 cm with 10 cm calibration, and 20 cm with 20 cm calibration.}\label{fig:IRaltcalib}
\end{subfigure}
\caption{IR$_{Cal}$ quantifying different distances using alternative calibrations.}
\label{fig:infraredError}
\end{figure*}

With IR$_{Cal}$, SparkFun:VL53L4CD, we considered two experimental setups: the sensor closest to the table and the sensor raised 3.7 cm above the table.
See Figure~\ref{fig:infrared}.
Figures~\ref{fig:infraredTouching} and~\ref{fig:infraredElevated} show the accuracy of measuring different distances with a 1 cm calibration for the two setups, respectively.
Note that the scale of the y-axis is logarithmic in Figure~\ref{fig:infraredTouching}.
The accuracy is highest when measuring distances of 1 or 2 cm.  The percentage error is higher with 10 and 20 cm.
With 20 cm measurements, the elevated setup provides a lower error.
Conceptually, this is due to the 3D cone shape of the sensor that is blocked by the table.
We anticipate the sensors to be mounted on the FLSs and the FLSs to fly at least a few centimeters above the ground.
Hence, the results of Figure~\ref{fig:infraredElevated} are most applicable. 

Similar to UWB, IR$_{Cal}$ measures a distance more accurately when it is calibrated for the distance. 
Figure~\ref{fig:IRaltcalib} shows this and an enhanced accuracy with longer distances.
The average percentage error is close to 0\% with 2, 10, and 20 cm.
We were not able to calibrate IR$_{Cal}$ to measure a 100 cm distance.

\subsection{IR$_{NoC}$, Sharp:GP2Y0A51SK0F}

\begin{figure*}
\centering
\begin{subfigure}[t]{0.47\linewidth}
  \centering
  \includegraphics[width=\linewidth]{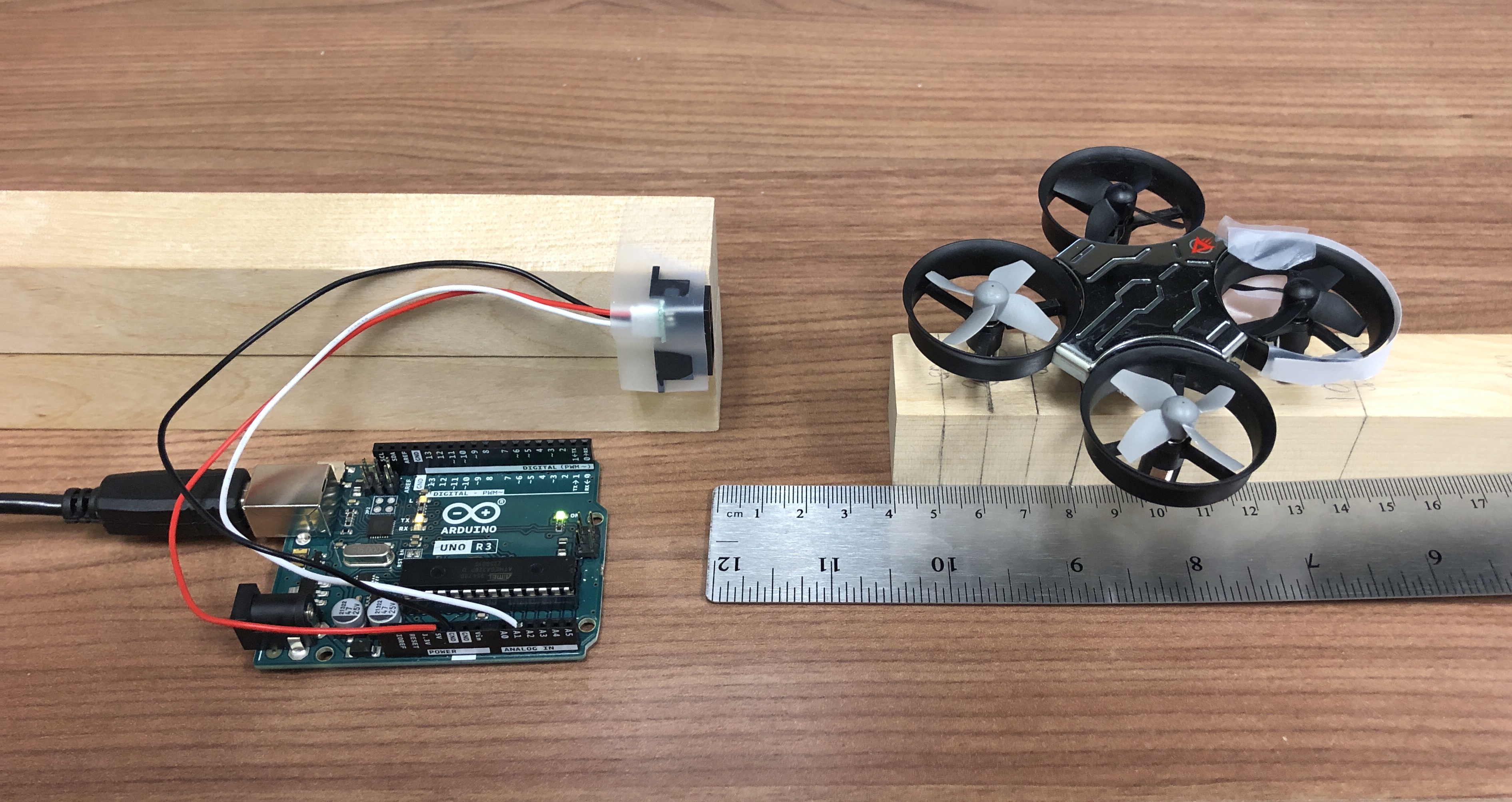}
  \caption{IR$_{NoC}$ Sharp:GP2Y0A51SK0F is elevated 3.7 cm.}\label{fig:sharpExp}
\end{subfigure}
\quad
\begin{subfigure}[t]{0.47\linewidth}
  \centering
  \includegraphics[width=\linewidth]{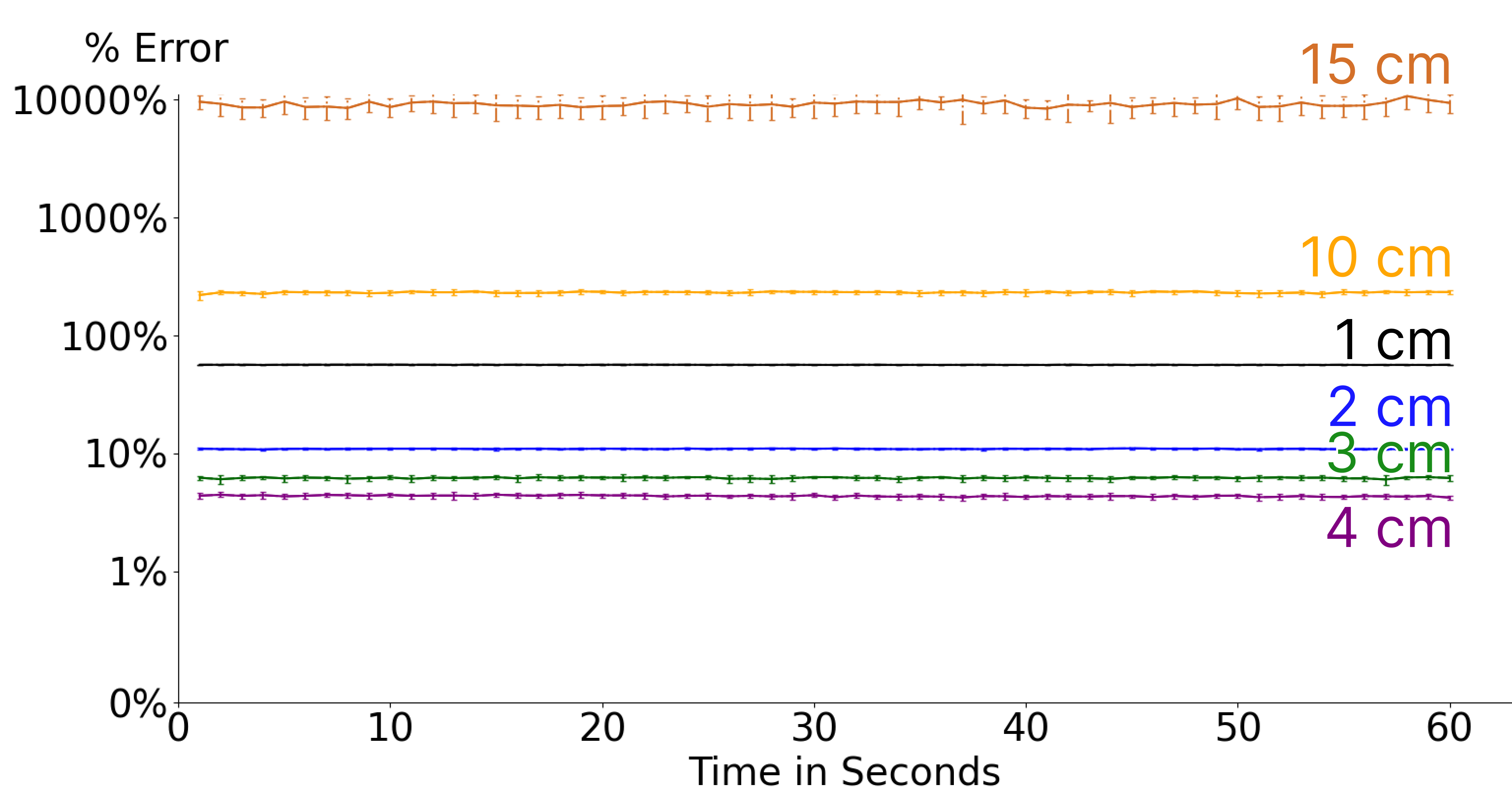}
  \caption{Percentage error measuring different distances using the IR$_{NoC}$ Sharp:GP2Y0A51SK0F sensor.}\label{fig:sharpError}
\end{subfigure}
\caption{IR$_{NoC}$ Sharp:GP2Y0A51SK0F.}
\label{fig:sharp}
\end{figure*}

\begin{figure*}
\centering
\begin{subfigure}[t]{0.47\linewidth}
  \centering
  \includegraphics[width=\linewidth]{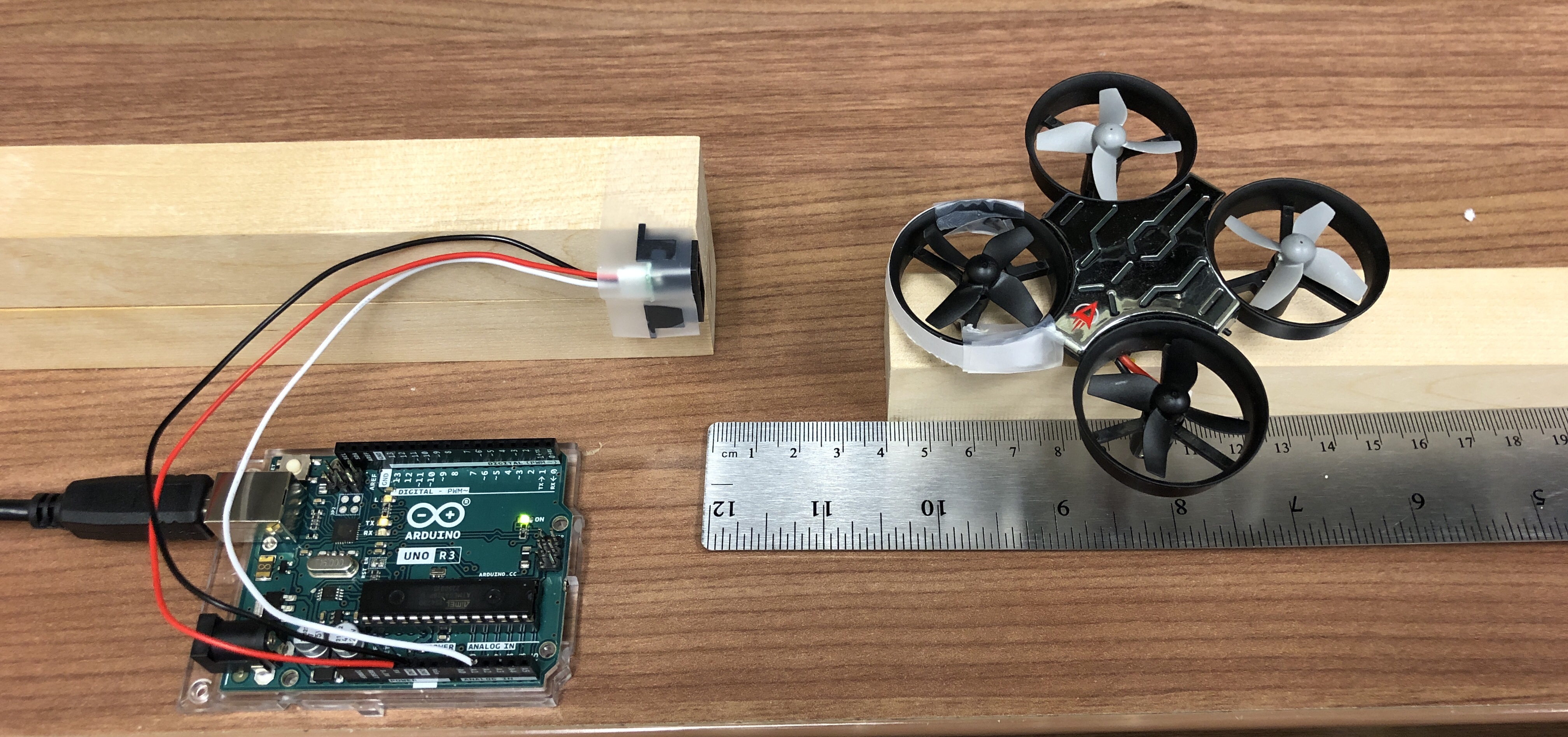}
  \caption{IR$_{NoC}$ Sharp:GP2Y0A51SK0F with a white piece of paper for the drone cage.}\label{fig:sharpWpaperExp}
\end{subfigure}
\quad
\begin{subfigure}[t]{0.47\linewidth}
  \centering
  \includegraphics[width=\linewidth]{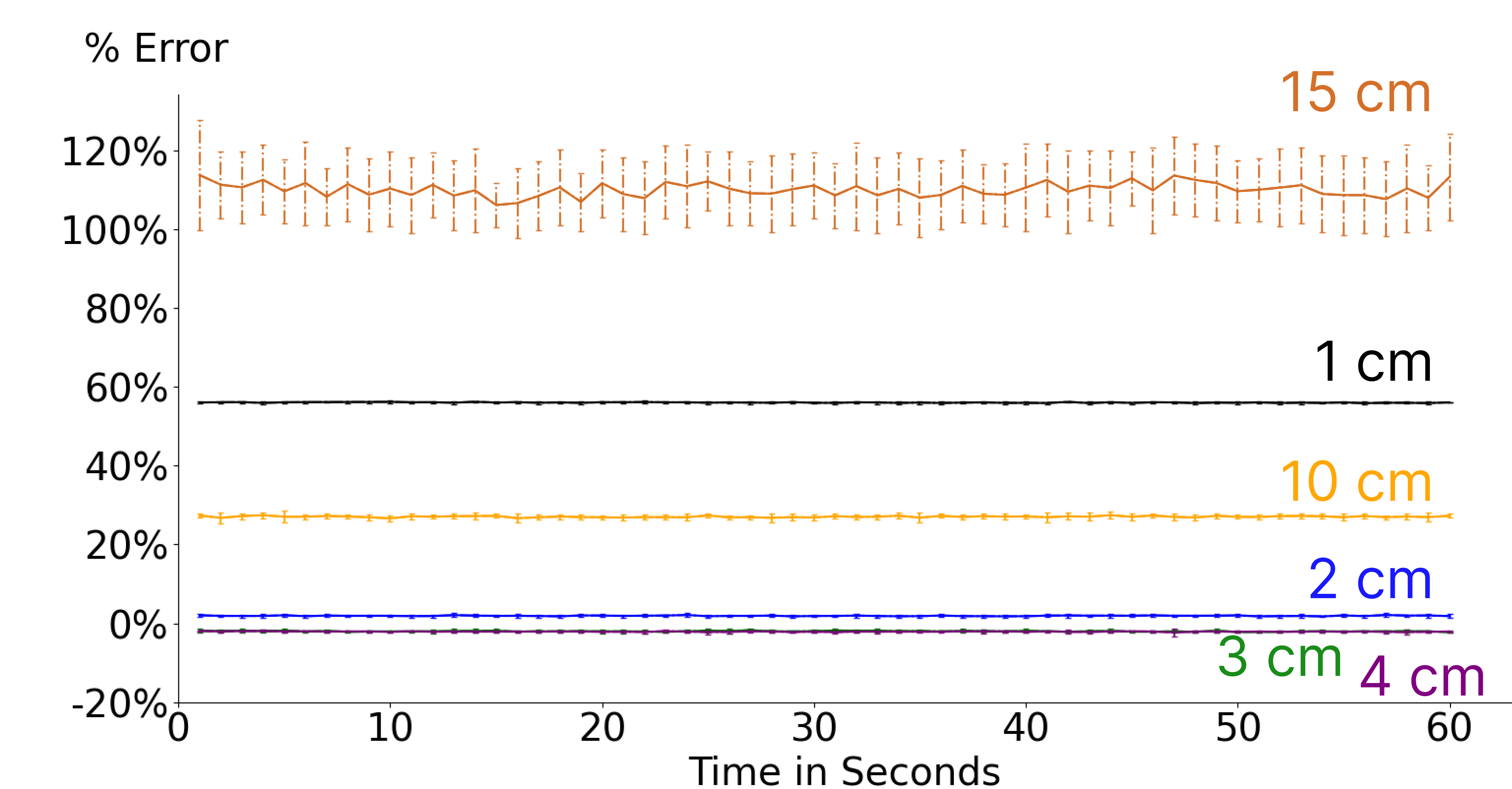}
  \caption{Percentage error measuring different distances using the experimental setup of Figure~\ref{fig:sharpWpaperExp}.}\label{fig:sharpWpaperError}
\end{subfigure}
\caption{IR$_{NoC}$ Sharp:GP2Y0A51SK0F.}
\label{fig:sharpWpaper}
\end{figure*}

IR$_{NoC}$, Sharp:GP2Y0A51SK0F, does not require calibration. 
It is more accurate when a piece of white paper is wrapped around the cages of drone blades, compare~\ref{fig:sharpExp} with~\ref{fig:sharpWpaperExp}.
Figures~\ref{fig:sharpError} and~\ref{fig:sharpWpaperError} show this sensor is highly accurate in measuring its minimum advertised distance of 2 cm, see Figure~\ref{fig:sharpError}.
Its accuracy is lower when measuring longer distances and worst at its maximum of 15 cm.
\section{Ultrasonic}\label{sec:us}
Similar to infrared sensing,
ultrasonic sensors work by emitting a high-frequency sound that bounces off of an object and then returns to be detected by the sensor. They are a type of time-of-flight sensor, meaning that distance is calculated by measuring the amount of time that the sound takes to return to the sensor:
\begin{equation}
    d = \frac{a \times t}{2}
\end{equation}
where $a$ is the speed\footnote{Speed of sound depends on the type of medium and the temperature of the medium.  Our experiments are conducted in room temperature.} of sound and $t$ is the recorded time. Ultrasonic sensors have a minimum and maximum distance that they can measure due to the measurement speed of the sensor and the dissipation of sound, respectively. If there are multiple objects in front of the ultrasonic sensor, the sound will bounce off the closest one, hence only the distance of the closest object will be returned. Typical ultrasonic sensors have a range of 2--400 cm with a resolution of 3 mm~\cite{mohammad2009using}.

\begin{figure*}[ht]
\centering
\begin{subfigure}{0.47\linewidth}
  \centering
  \includegraphics[width=\linewidth]{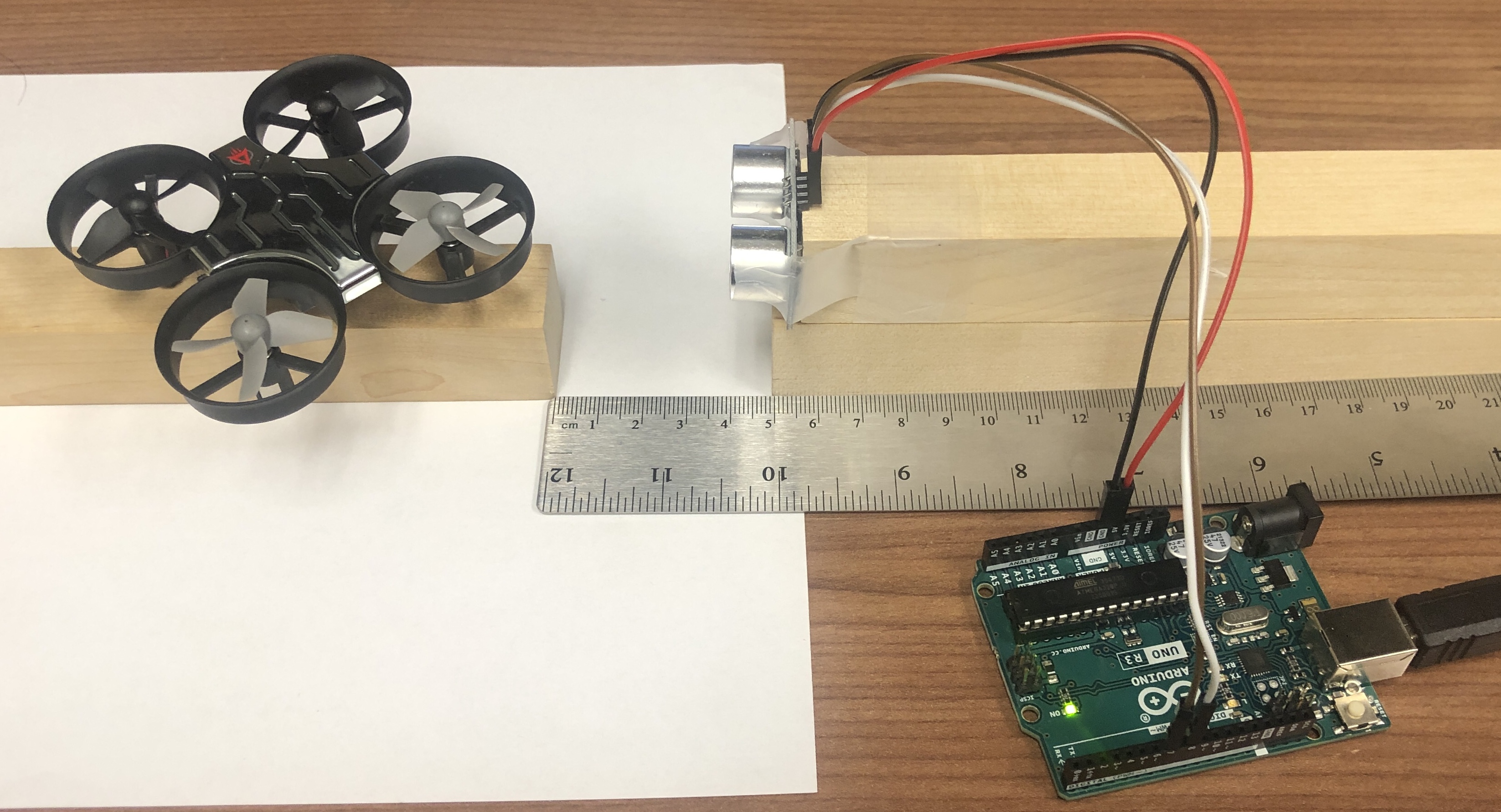}
  \caption{Ultrasonic is elevated 3.7 cm.}\label{fig:usExp}
\end{subfigure}
\quad
\begin{subfigure}{0.47\linewidth}
  \centering
  \includegraphics[width=\linewidth]{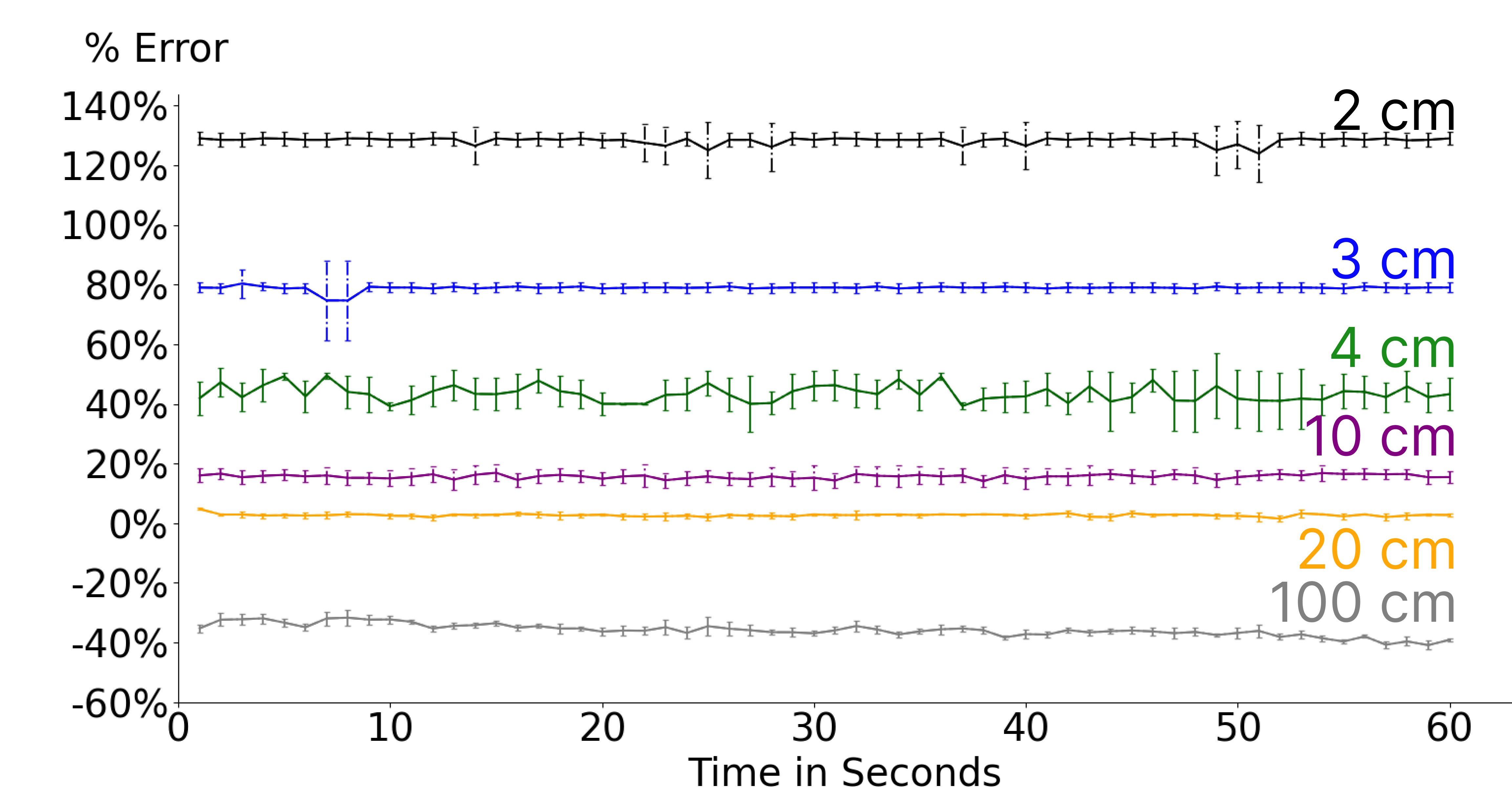}
  \caption{Percentage error measuring different distances using the ultrasonic sensor.}\label{fig:usError}
\end{subfigure}
\caption{Ultrasonic.}
\label{fig:ultrasonic}
\end{figure*}

The size of an object that can be detected by an ultrasonic sensor is limited by the sounds' wavelength. An ultrasonic sensor that emits a 40 kHz burst of sound has a wavelength of 8.6 mm. Objects smaller than this wavelength will cause scattering of the sound, but will not cause the sound to bounce back and return to the sensor. Therefore, the sensor may not detect objects smaller than its wavelength. 

The sound that is emitted from an ultrasonic sensor spreads the further it travels away from the sensor, similar to the beam of a flashlight. The cone-shaped emission of sound results in a set angular resolution for an ultrasonic sensor. This field of view (FoV) is typically around 30$^{\circ}$, and determines the spatial resolution of an FLS. 
To detect FLSs in more than this limited angular area, an FLS may use (1) a single ultrasonic sensor that rotates, (2) multiple ultrasonic sensors that are oriented around the FLS, or (3) communicate with other FLSs surrounding it to gather their detected FLSs.

 When using ultrasonic sensors to measure distance, it is best practice to ensure the surface properties of all objects are textured~\cite{Kleeman2008}.
 Ultrasonic sensors have a tendency towards specular reflections when looking at shiny, mirrored, or glass objects. With specular reflections, the angle of reflection is equal to the angle of incidence, meaning that the sound is less likely to return to the sensor. 
 Thus, shiny objects may not be detected or may appear far away.
 These considerations are important when designing FLSs.

Since ultrasonic sensors work only by measuring the amount of time that elapses between when a sound is emitted and when it is detected, there can be issues in a system consisting of multiple FLSs using ultrasonic sensors. For example, if two FLSs are 1 m away from each other and are trying to measure that distance, if their ultrasonic sensors are pointed at each other it is likely that they could read a distance of only 0.5 m due to interference between the two sensors. Therefore, with an FLS display, more advanced signal processing and sensors would be required, such as amplitude- or frequency-modulated ultrasonic sensors.

Figure~\ref{fig:usError} shows the percentage error observed in measuring different distances with ultrasonic~\cite{hcsr04}.  
The experimental setup is shown in Figure~\ref{fig:usExp}.
The ultrasonic is most accurate for measuring distances of 10 and 20 cm.
Its accuracy is lower with shorter ($\leq$ 4 cm) and longer distances ($\geq$ 100 cm). 

\begin{figure}
\centering
\includegraphics[width=0.9\columnwidth]{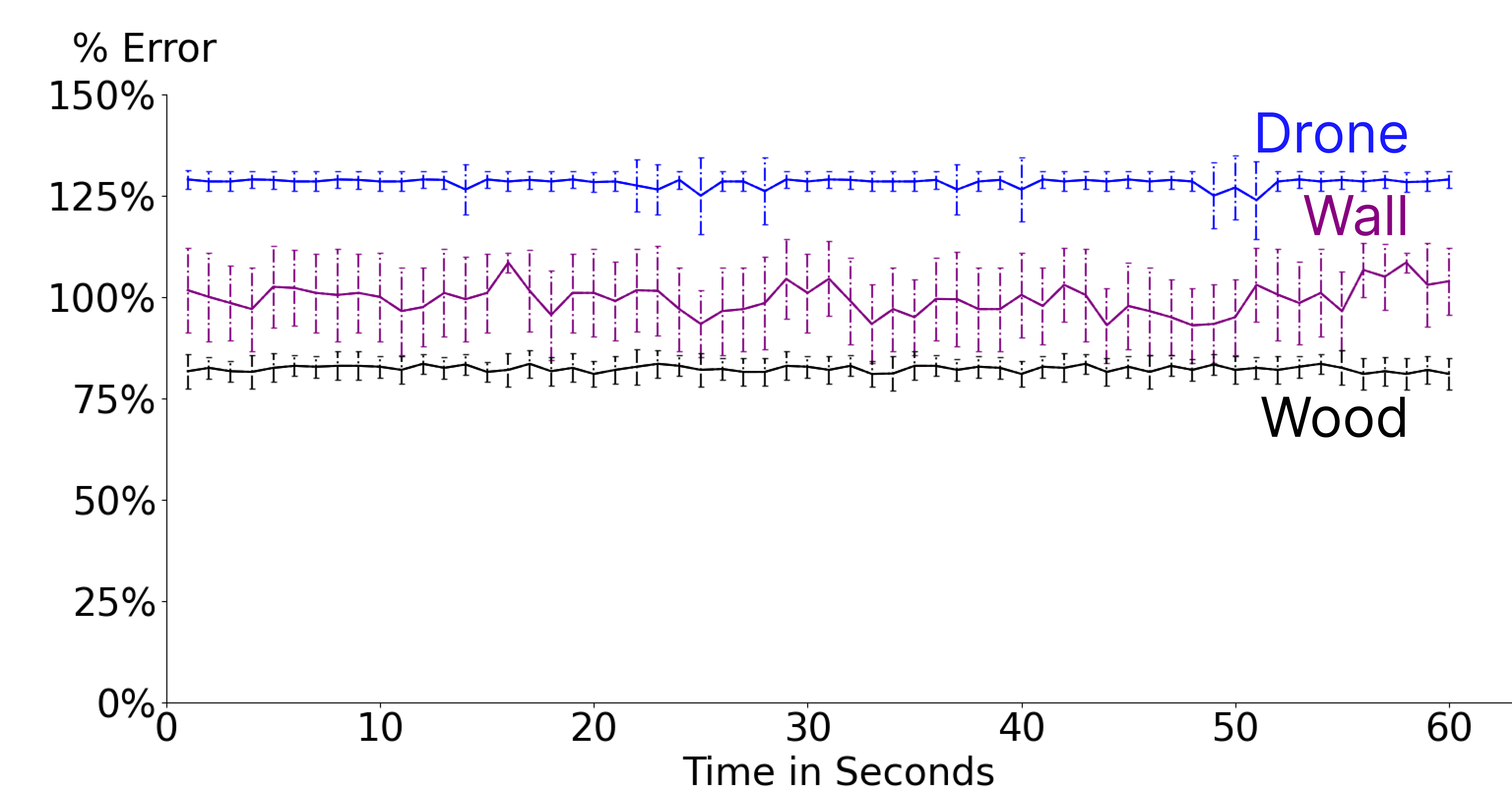}\hfill
\caption{Ultrasonic with different material for a target object 2 cm away.}
\label{fig:composition}
\end{figure}

The error observed when measuring small distances (2 cm) is sensitive to the material used by the object, see Figure~\ref{fig:composition}.

\begin{figure*}
\centering
\begin{subfigure}[t]{0.31\linewidth}
  \centering
  \includegraphics[width=\linewidth]{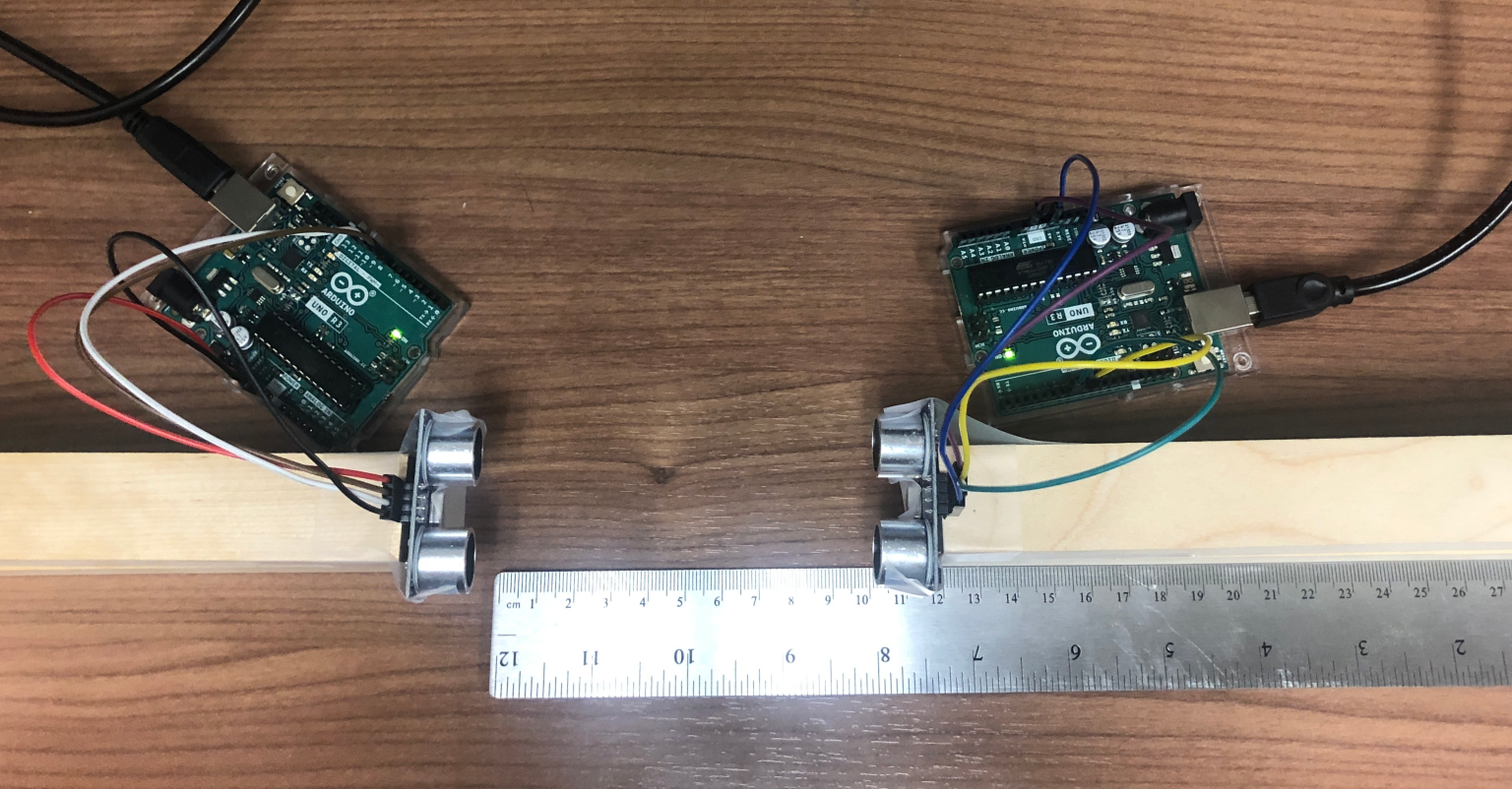}
  \caption{Experimental setup.}\label{fig:usFacingExp}
\end{subfigure}
\quad
\begin{subfigure}[t]{0.31\linewidth}
  \centering
  \includegraphics[width=\linewidth]{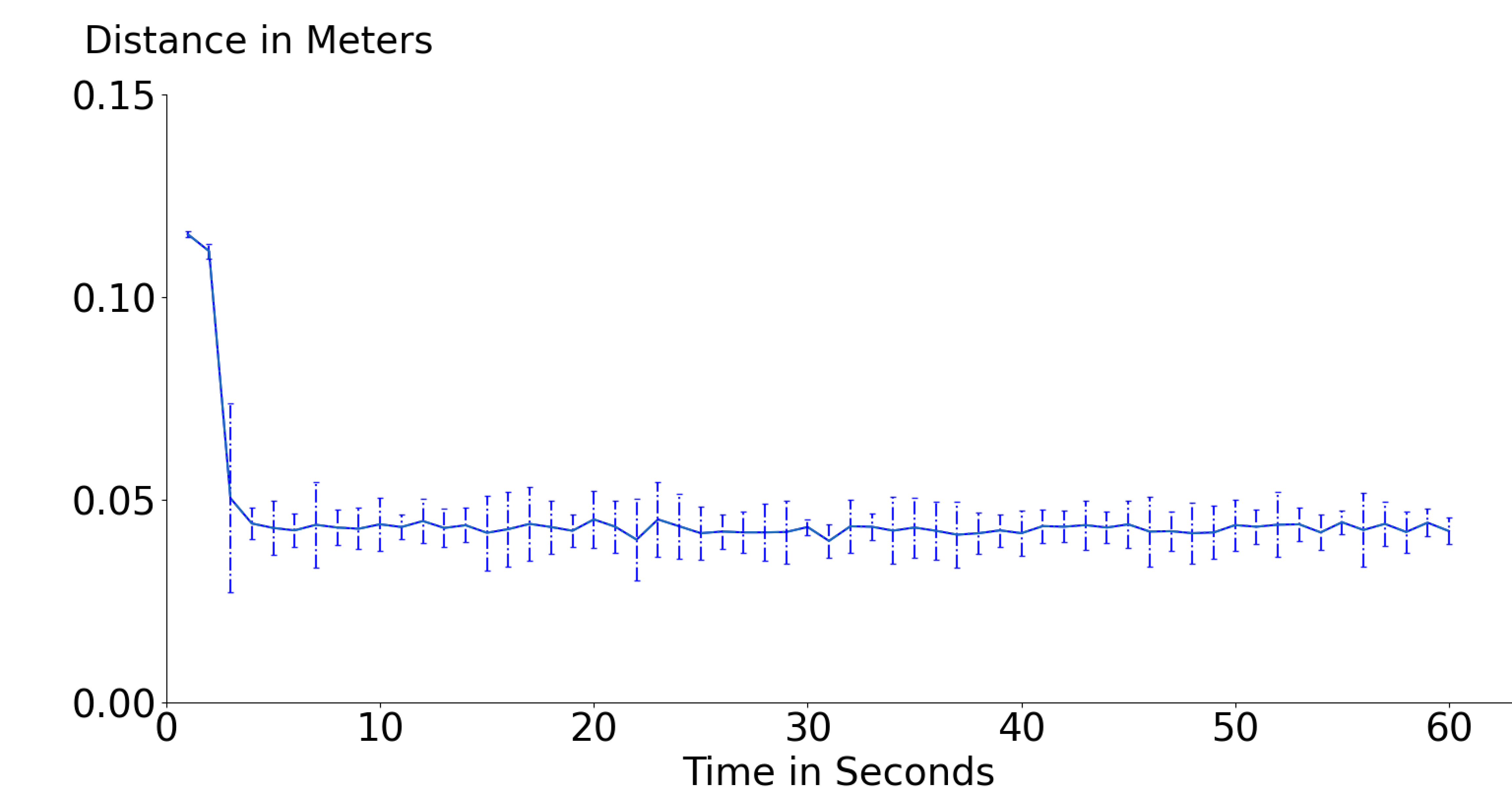}
  \caption{Measured distance when 10 cm apart.}\label{fig:usFacingDist}
\end{subfigure}
\quad
\begin{subfigure}[t]{0.31\linewidth}
  \centering
  \includegraphics[width=\linewidth]{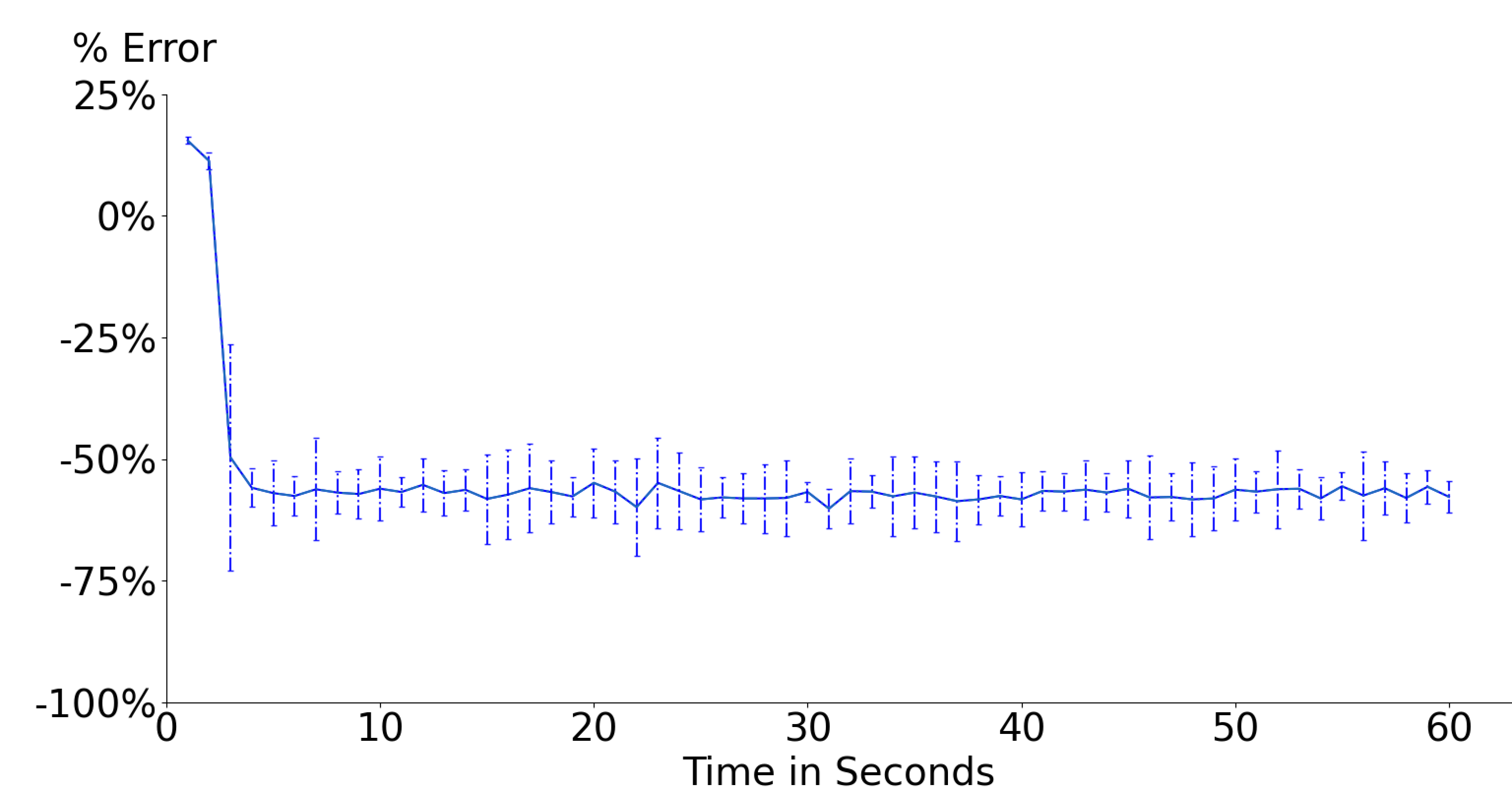}
  \caption{A 50\% error in measuring distance is anticipated with two sensors interfering with one another.}\label{fig:usFacingError}
\end{subfigure}
\caption{Two ultrasonic sensors facing one another.}
\label{fig:facingultrasonic}
\end{figure*}

Figure~\ref{fig:facingultrasonic} shows an experiment with two Ultrasonic sensors facing one another and 10 cm apart.  The sensors interfere with one another with each reporting a distance of 5 cm.

\begin{figure*}
\centering
\begin{subfigure}[t]{0.31\linewidth}
  \centering
  \includegraphics[width=\linewidth]{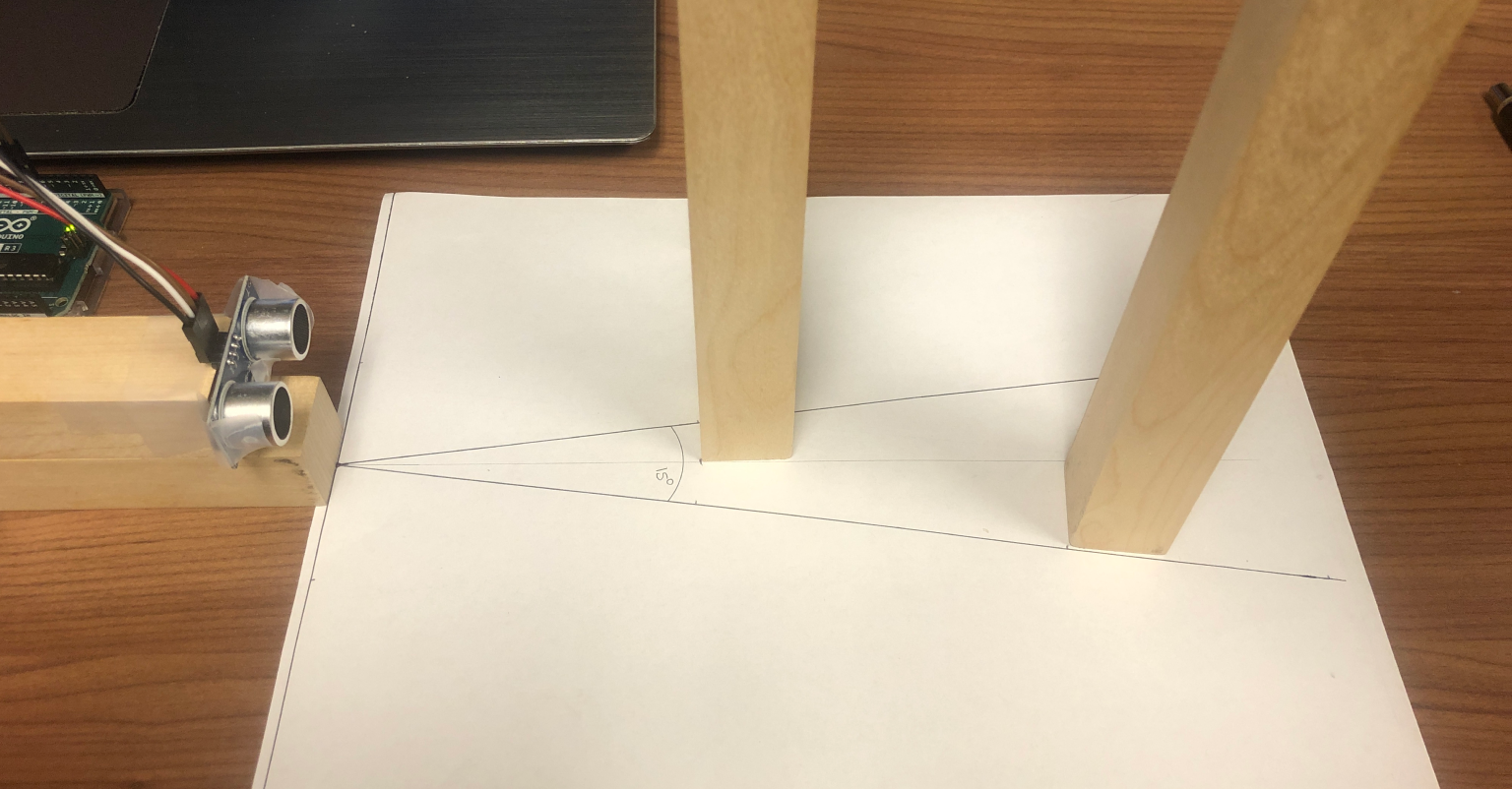}
  \caption{Experimental setup.}\label{fig:us2ObjExp}
\end{subfigure}
\quad
\begin{subfigure}[t]{0.31\linewidth}
  \centering
  \includegraphics[width=\linewidth]{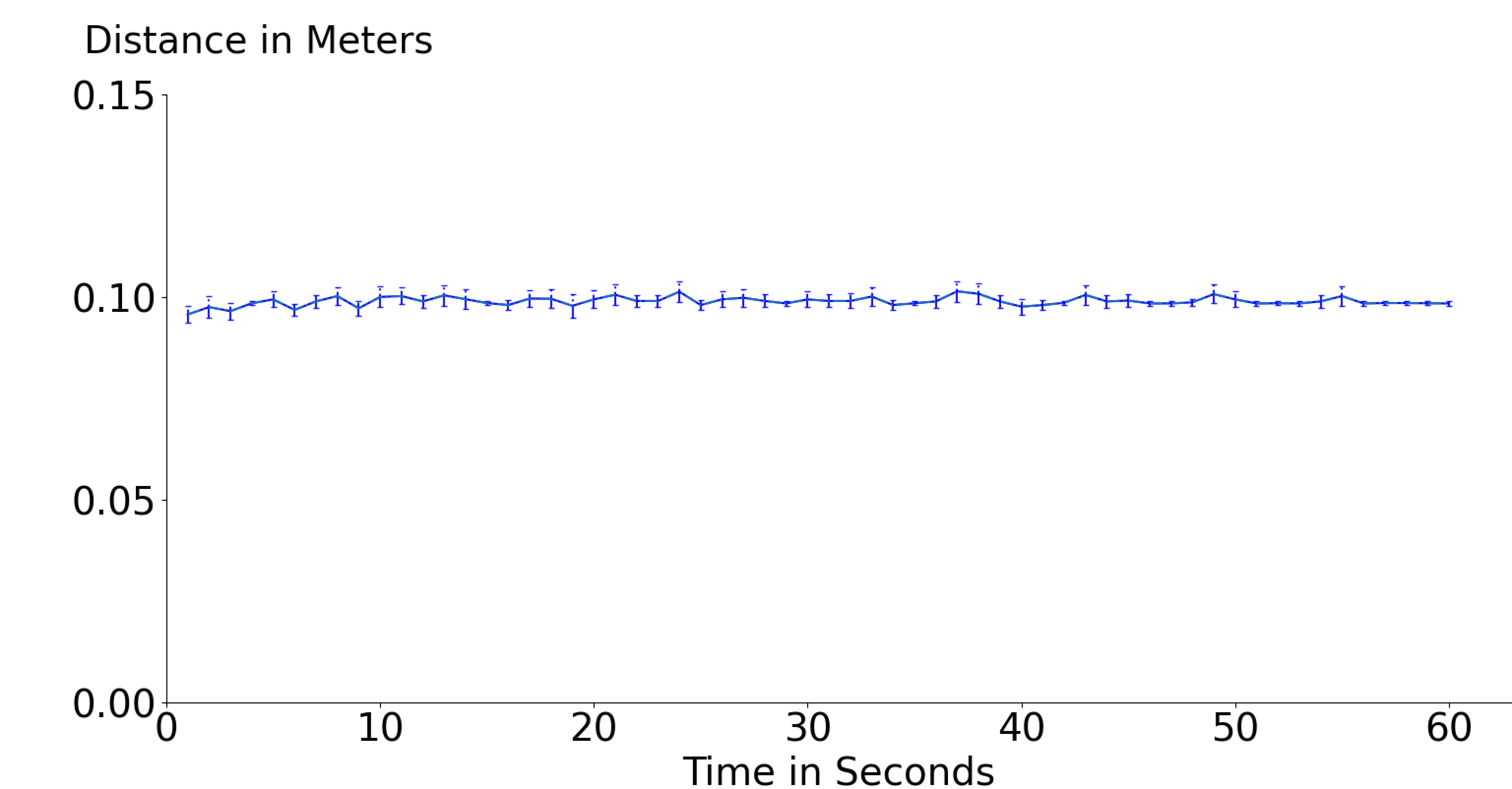}
  \caption{Sensor reports distance to the closest object.}\label{fig:us2ObjDist}
\end{subfigure}
\quad
\begin{subfigure}[t]{0.31\linewidth}
  \centering
  \includegraphics[width=\linewidth]{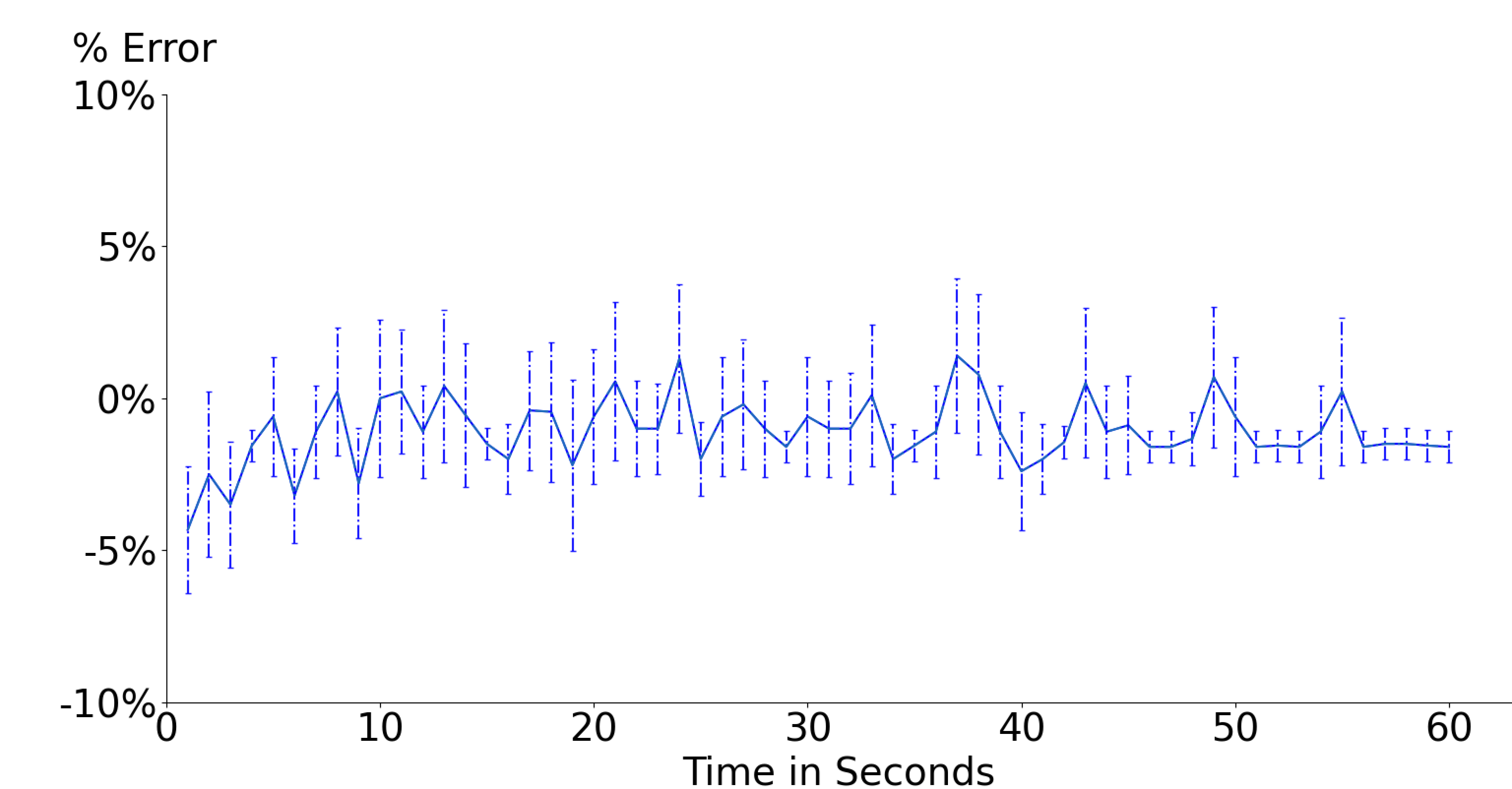}
  \caption{Percentage error in measuring the distance to closest object, 10cm away.}\label{fig:us2ObjError}
\end{subfigure}
\caption{Ultrasonic sensor with two objects, 10 and 20 cm away from the sensor.}
\label{fig:facingultrasonic2objs}
\end{figure*}

Figure~\ref{fig:facingultrasonic2objs} shows the Ultrasonic sensor with two wooden square pillars in its field of view.
One is 10 cm and the other is 20 cm away from the sensor.
The sensor reports the distance to the closest object.

\section{Conclusions and Future Research Directions}\label{sec:conc}
This paper compares several time-of-flight sensors to measure small distances.
The sensors are based on different technologies:  radio (UWB), light (IR), and sound (US).  
Only one product, IR$_{Cal}$ (SparkFun:VL53L4CD), measured distances as small as 1 cm with a high accuracy.
Its accuracy depends on the distance used for its calibration with a calibration of 1 cm providing the highest accuracy, 10-15\%.

The obtained results motivate techniques that fuse measurements from different sensors to estimate distance between FLSs.  
Examples include Inertial-Measurement-Unit (IMU) with visual features from a camera~\cite{orb2017,vins2018,openvins2020}, LiDAR with odometry~\cite{lidar2017,loam2014,floam2021,loamlivox2020}, UWB with IMU and images from a camera~\cite{xu2021,xu2022,viunet2023,uwbIMUvisual2021}, UWB with IMU and LiDAR~\cite{liro2021}.

We are extending this study in several directions.  
First, we are considering sensors that use Frequency-Modulated Continuous Wave~\cite{fmcw2013} (FMCW) instead of time-of-flight.
An example is radar~\cite{radar2018}.
Our objective is to identify the least expensive and most accurate sensor to measure small distances ($\leq$ 1cm).
We will equip FLSs with the identified sensor to design and implement localization, collision avoidance, and haptic interaction techniques.
An FLS display will illuminate objects and avatars of a metaverse, enabling users to see them without wearing head mounted displays and have haptic interactions without wearing gloves.

\section{Acknowledgments}
This research was supported in part by the NSF grant IIS-2232382.

\bibliography{refs}
\end{document}